\documentclass[twocolumn]{aastex62}

\received{}
\revised{}
\accepted{}
\submitjournal{ApJ}

\shorttitle{H$\alpha$ emitters with the $\nu^2$GC}
\shortauthors{Ogura et al.}

\begin{document}

\title{Quantifying the effect of field variance on the H$\alpha$ luminosity function with the New Numerical Galaxy Catalogue ($\nu^2$GC)}


\author{Kazuyuki Ogura}
\affil{Faculty of Education, Bunkyo University, \\
3337 Minamiogishima, Koshigaya, Saitama, 343-8511, Japan}
\email{ogurakz@koshigaya.bunkyo.ac.jp}

\author{Masahiro Nagashima}
\affil{Faculty of Education, Bunkyo University, \\
3337 Minamiogishima, Koshigaya, Saitama, 343-8511, Japan}

\author{Rhythm Shimakawa}
\affil{Subaru Telescope, National Astronomical Observatory of Japan, \\
650 North A’ohoku Place, Hilo, HI, 96720, USA}

\author{Masao Hayashi}
\affil{National Astronomical Observatory of Japan, \\
2-21-1 Osawa,Mitaka, Tokyo 181-8588, Japan}

\author{Masakazu A. R. Kobayashi}
\affil{Faculty of Natural Sciences, National Institute of Technology, Kure College, \\
2-2-11, Agaminami, Kure, Hiroshima, 737-8506, Japan}

\author{Taira Oogi}
\affil{Kavli Institute for the Physics and Mathematics of the Universe, \\
The University of Tokyo Institutes for Advanced Study, the University of Tokyo, \\ 
5-1-5, Kashiwanoha, Kashiwa, 277-8583, Japan}

\author{Tomoaki Ishiyama}
\affil{Institute of Management and Information Technologies, Chiba University, \\
1-33, Yayoi-cho, Inage-ku, Chiba 263-8522, Japan}

\author{Yusei Koyama}
\affil{Subaru Telescope, National Astronomical Observatory of Japan, \\
650 North A’ohoku Place, Hilo, HI, 96720, USA}

\author{Ryu Makiya}
\affil{Kavli Institute for the Physics and Mathematics of the Universe, \\
The University of Tokyo Institutes for Advanced Study, the University of Tokyo, \\ 
5-1-5, Kashiwanoha, Kashiwa, 277-8583, Japan}

\author{Katsuya Okoshi}
\affil{Tokyo University of Science, \\
102-1 Tomino, Oshamambe-cho, Yamakoshi-gun, Hokkaido, 049-3514, Japan }

\author{Masato Onodera}
\affil{Subaru Telescope, National Astronomical Observatory of Japan, \\
650 North A’ohoku Place, Hilo, HI, 96720, USA}
\affil{Department of Astronomical Science, The Graduate University for Advanced Studies, SOKENDAI, \\
Mitaka Tokyo 181-8588, Japan}

\author{Hikari Shirakata}
\affil{Department of Cosmosciences, Hokkaido University, \\
N10 W8, Kitaku, Sapporo, 060-0810, Japan}

\begin{abstract} 
We construct a model of H$\alpha$ emitters (HAEs) based on a semi-analytic galaxy formation model, the New Numerical Galaxy 
Catalog ($\nu^2$GC). In this paper, we report our estimate for the field variance of the HAE distribution. By calculating the H$\alpha$ luminosity 
from the star-formation rate of galaxies, our model well reproduces the observed H$\alpha$ luminosity function (LF) at $z=0.4$. 
The large volume of the $\nu^2$GC makes it possible to examine the spatial distribution of HAEs over a region of (411.8 Mpc)$^3$ 
in the comoving scale. The surface number density of $z=0.4$ HAEs with $L_{\rm H\alpha} \geq 10^{40}$~erg~s$^{-1}$ is 308.9~deg$^{-2}$. 
We have confirmed that the HAE is a useful tracer for the large-scale structure of the Universe because of their significant overdensity 
($>5\sigma$) at clusters and the filamentary structures. The H$\alpha$ LFs within a survey area of $\sim$2~deg$^2$ (typical for previous
observational studies) show a significant field variance up to $\sim$1~dex. 
Based on our model, one can estimate the variance on the H$\alpha$ LFs within given survey areas.
\end{abstract}

\keywords{galaxies: clusters: general --- galaxies: formation --- galaxies: evolution --- galaxies: luminosity function --- methods: numerical}


\section{Introduction} \label{sec:intro} 

So far, various extensive observations of star-forming galaxies have shown that galaxies do not distribute uniformly 
and that their spatial distribution shows various structures at various spatial scales 
\citep[e.g.,][]{1986ApJ...302L...1D, 2012ApJ...761...14H, 2012MNRAS.424..564R}. 
The formation of such a large-scale structure (LSS) of the Universe has been investigated by $N$-body simulations. 
Structure formation is powered by gravitational instabilities of cold dark matter (CDM) and form dark matter halos 
where galaxies form and evolve \citep[e.g.,][]{2005Natur.435..629S}.

For investigating the evolution of the galaxies and structure formation, emission-line galaxies (ELGs) have been often used
\citep[e.g.,][]{2003ApJ...586L.111S, 2003ApJ...582...60O, 2004AJ....128.2073H, 2004AJ....128..569M, 2008ApJS..176..301O, 
2008MNRAS.388.1473G, 2010MNRAS.402.1980H, 2012ApJ...759..133M, 2013IAUS..295...74K, 2014ApJ...796...51D, 
2015MNRAS.452.3948K, 2017MNRAS.468L..21S, 2017PASJ...69...51O, 2018PASJ...70S..13O, 2018PASJ...70S..14S}
because emission lines from galaxies, such as [O~{\sc ii}] ($\lambda\lambda$3727, 3730), [O~{\sc iii}] 
($\lambda\lambda$4959, 5007), and H$\alpha$ ($\lambda$6563), are good indicators for the star-formation rate (SFR) of 
galaxies \citep[e.g.,][]{1998ARA&A..36..189K, 2004AJ....127.2002K, 2006ApJ...642..775M, 2013MNRAS.430.1042H, 
2016MNRAS.462..181S}.
A strong point to focus on ELGs is that we 
can observe them over a significantly wide area by utilizing the combination of narrow-band (NB) filters and wide-field 
cameras.
Note that, although spectroscopic surveys including the integral field spectroscopy are more powerful than NB 
surveys to identify emission lines and to measure accurate line luminosities, 
the survey area with enough depth is limited. Therefore, it is difficult to trace the LSS by spectroscopic observations,
except for some samples in the bright regime \citep[e.g.,][]{2015A&A...583A.128D}.

Among such ELGs, H$\alpha$ emitters (HAEs) are often focused since H$\alpha$ emission is well calibrated and only 
mildly affected by the dust attenuation compared to other emission lines at wavelengths bluer than H$\alpha$
\citep[e.g.,][]{2010MNRAS.402.2017G, 2010MNRAS.409..421G, 2013MNRAS.436.1130S, 2015MNRAS.451.2303S}. 
Moreover, the H$\alpha$ emission line from galaxies at a wide redshift range ($0 \leq z \lesssim 2.6$) can be observed with 
NB imaging using ground-based optical and NIR instruments. So far, many surveys using NB filters have been conducted 
to make a large sample of HAEs over a wide redshift range \citep[e.g.,][]{2004MNRAS.354.1103K, 2007ApJ...657..738L, 
2011ApJ...726..109L, 2011MNRAS.416.2041M, 2011MNRAS.415.2993H, 2011PASJ...63S.415T, 2011ApJ...734...66K, 
2013MNRAS.428.1551K, 2011MNRAS.411..675S, 2013MNRAS.428.1128S, 2015MNRAS.451.2303S, 
2013MNRAS.433..796D, 2017MNRAS.471..629M, 2017MNRAS.465.2916S, 2018PASJ...70S..17H}.

The field variance (or ``cosmic variance'') has been recognized as a serious problem in observational studies of high-$z$ 
galaxies \citep{2004ApJ...600L.171S, 2008ApJ...676..767T}. It is the uncertainty in measuring the number density of 
galaxies due to underlying density fluctuation of dark matter.
Indeed, while luminosity functions (LFs) of ELGs obtained by some different surveys generally show good agreement, 
some of them are in some disagreement by a factor of a few \citep[e.g.,][]{2011ApJ...726..109L, 2012PASP..124..782L, 
2012MNRAS.420.1926S, 2013ApJ...779...34C, 2014MNRAS.438.1377S}. 
Because most of previous surveys covered less than 2~deg$^2$ area, the results could be caused by the field variance. 
To reduce the effect of the field variance, \cite{2015MNRAS.451.2303S} performed $\sim$10~deg$^2$ survey while the survey 
depth is somewhat shallow. Very recently, \cite{2018PASJ...70S..17H} construct the largest sample of ELGs so far by utilizing 
deep and wide data in a $\sim$16~deg$^2$ area provided by the first public data release \citep[][]{2018PASJ...70S...8A} of 
the Subaru Strategic Survey with Hyper Suprime-Cam \citep{2012SPIE.8446E..0ZM} on the Subaru telescope 
\citep[HSC-SSP;][]{2018PASJ...70S...4A}. 
The H$\alpha$ LF at $z=0.4$ derived by \cite{2018PASJ...70S..17H} 
shows a good agreement with previous surveys reported by \cite{2007ApJ...657..738L}, \cite{2013MNRAS.433..796D} and 
\cite{2013MNRAS.428.1128S}.
However, all of these studies employ different threshold of equivalent width (EW) for selecting 
HAEs; specifically, the rest-frame equivalent width ($EW_0$) $\geq 40$~{\AA} \citep[][]{2018PASJ...70S..17H}, 
$EW_0 \geq 11$~{\AA} \citep[][]{2007ApJ...657..738L}, $EW_0 \geq 100$ {\AA} \citep[][]{2013MNRAS.433..796D}, and 
$EW_0 \geq 25$~{\AA} \citep[][]{2013MNRAS.428.1128S}. Furthermore, \cite{2018PASJ...70S..17H} show that H$\alpha$ 
LFs obtained in different fields show significant variance up to $\sim$1~dex. 
How large is the field variance in a given survey area ? 
How wide is survey area required to converge the LF ? These are still open questions.

To tackle this issue, we examine the field variance of the spatial distribution of galaxies by using a semi-analytic galaxy 
formation model, the New Numerical Galaxy Catalog \citep[$\nu^2$GC;][]{2016PASJ...68...25M,  2019MNRAS.482.4846S}. 
A remarkable aspect of the $\nu^2$GC is a large comoving volume up to $\sim$4.5~Gpc$^3$ with sufficient mass resolution 
based on the state-of-the-art cosmological $N$-body simulations by \cite{2015PASJ...67...61I}. This enables us to examine 
various statistical properties of galaxies over a wide area. Indeed, the $\nu^2$GC is successful to reproduce observed 
statistical properties of galaxies in a wide redshift range of $0\leq z<6$ such as the H{\sc i} mass function, broad-band 
LF, and cosmic star-formation history \citep[see][for more details]{2016PASJ...68...25M, 
2019MNRAS.482.4846S}. By utilizing the $\nu^2$GC, we construct a model of HAEs to examine the field variance of 
their spatial distribution. In this paper, we focus on HAEs at $z=0.4$ at which extensive observations of HAEs has been 
conducted \citep{2007ApJ...657..738L, 2013MNRAS.433..796D, 2013MNRAS.428.1128S, 2018PASJ...70S..17H}. 

This paper is organized as follows. We present overview of the $\nu^2$GC and the HAE model in Section \ref{sec:model}. 
The properties of model HAEs including the H$\alpha$ LF are described in Section \ref{sec:HAEprop}. 
In Section \ref{sec:discussion}, we discuss the field variance on the H$\alpha$ LF. We then give our conclusion in 
Section \ref{sec:conclud}. 
Throughout this paper, we employ a $\Lambda$CDM cosmology in which $h_0=0.68$, $\Omega_{\rm M}=0.31$, and  
$\Omega_{\rm \Lambda}=0.69$ \citep{2014A&A...571A..16P}, unless otherwise stated. Given these parameters, 1~arcsec 
corresponds to 5.515~physical~kpc at $z=0.4$. 
We adopt the Chabrier initial mass function (IMF) in the mass range of 0.1 -- 100~$M_{\odot}$ \citep{2003PASP..115..763C}.
All magnitudes are given in the AB system \citep[][]{1983ApJ...266..713O}.

\section{Model} \label{sec:model}  

\subsection{The $\nu^2$GC} \label{sec:nu2gc} 

The $\nu^2$GC \citep{2016PASJ...68...25M, 2019MNRAS.482.4846S} is a semi-analytic model for the galaxy formation, 
which is an updated version of the Numerical Galaxy Catalog \citep[$\nu$GC;][]{2005ApJ...634...26N, 2014ApJ...794...69E, 
2015MNRAS.450L...6S}. We use the $\nu^2$GC-S\footnote{\url{http://hpc.imit.chiba-u.jp/~ishiymtm/db.html}} 
\citep[][]{2015PASJ...67...61I}, which is an $N$-body simulation with a 
box of (280 $h^{-1}$~Mpc)$^3$ comoving scale [corresponding to (411.8~Mpc)$^3$ when we adopt $h_0=0.68$] containing 
2048$^3$ dark matter particles, for constructing merger trees of dark matter halos. The particle mass resolution, the minimum 
halo mass and the total number of halos in the box are 2.20$\times10^8$~$M_{\odot}$, 
8.79$\times10^9$~$M_{\odot}$ and 6,575,486, respectively. The large comoving volume of this simulation enables us to examine 
the statistical properties of galaxies including the spatial distribution over a wide area. Note that 411.8~Mpc in the comoving 
scale at $z=0.4$ corresponds to 14.8~degree on the sky.

The $\nu^2$GC includes many physical processes involved in the galaxy formation and evolution. We briefly summarize here the 
baryonic evolution model in the $\nu^2$GC. See \cite{2016PASJ...68...25M} and \cite{2019MNRAS.482.4846S} for further details 
of the model. We assume that a dark matter halo is filled with the hot gas with the virial temperature. The hot gas cools through the 
radiation cooling to form a gas disk. Here, we employ a scheme of the gas cooling rate proposed by \cite{1991ApJ...379...52W} and 
a cooling function provided by \cite{1993ApJS...88..253S}. The cold gas in the disk condenses to form stars. The star-formation rate
(SFR) is given by 
\begin{equation}
SFR = M_{\rm cold} / \tau_{\rm star},
\end{equation} 
where $M_{\rm cold}$ and $\tau_{\rm star}$ are the cold gas mass and the star-formation timescale, respectively. 
By assuming that the star-formation in the galactic disk is related to the dynamical time scale of the disk, 
$\tau_{\rm d}$ ($\equiv r_{\rm d} / V_{\rm d}$, where $r_{\rm d}$ and $V_{\rm d}$ are the radius of the 
disk and the disk rotation velocity, respectively), the star-formation timescale is given by the following formula:
\begin{equation}
\tau_{\rm star} = \varepsilon_{\rm star}^{-1} \tau_{\rm d} \left[1 + \Bigl(\frac{V_{\rm d}}{V_{\rm hot}} \Bigr)^{\alpha_{\rm star}} \right],
\end{equation}
where $\alpha_{\rm star}$, $\varepsilon_{\rm star}$, and $V_{\rm hot}$ are free parameters 
that are determined by fitting observed LFs ($r$- and $K$-bands) and cold neutral gas mass function at $z=0$
based on a Markov chain Monte Carlo (MCMC) method. The best fit values of these parameters are $\alpha_{\rm star} = -2.14$,
$\varepsilon_{\rm star} = 0.48$, and $V_{\rm hot} = 121.64$~km~s$^{-1}$
\citep[see][for the detail of the parameter tuning]{2016PASJ...68...25M, 2019MNRAS.482.4846S}.

As for the feedback, we assume that Type-II supernovae reheat a part of the cold gas to reject it from the galaxies at 
a rate of $M_{\rm cold} / \tau_{\rm reheat}$ (the effects of the type-Ia supernovae are negrected).
The reheating timescale $\tau_{\rm reheat}$ is given by following equation:
\begin{equation}
\tau_{\rm reheat} = \frac{\tau_{\rm star}}{\beta(V_{\rm d})},
\end{equation}
where 
\begin{equation}
\beta(V_{\rm d}) \equiv \Bigl(\frac{V_{\rm d}}{V_{\rm hot}}\Bigr)^{-\alpha_{\rm hot}},
\end{equation}
with free parameters of $V_{\rm hot}$ 
[the same as $V_{\rm hot}$ in equation (2), with a best fit value of 121.64~km~s$^{-1}$]
and $\alpha_{\rm hot}$ (the best fit value is $\alpha_{\rm hot}$ = 3.92) is the reheating timescale.

When two or more galaxies merge together, we assume that a starburst occurs and consume the cold gas in the bulge. 
The mass of stars formed by a starburst, $\Delta M_{\rm star, burst}$, is given by
\begin{equation}
\Delta M_{\rm star, burst} = \frac{\alpha}{\alpha + \beta + f_{\rm BH}} M^{0}_{\rm cold},
\end{equation}
where $M^{0}_{\rm cold}$ is the cold gas mass right after the burst, $\alpha$ is the locked-up mass fraction,
$\beta$ is defined by the equation (4), and $f_{\rm BH}$ is the fraction of the gas which is accreted onto
the supermassive black hole (SMBH). 
The locked-up mass fraction $\alpha$ is set to be consistent with IMF ($\alpha=0.52$ for the Chabrier IMF).
We set $f_{\rm BH}$ to reproduce observed relation between host
bulge mass and SMBH mass  \citep[$f_{\rm BH}=0.02$, see][]{2019MNRAS.482.4846S}.
Note that even in the case of the minor merger, a starburst occurs. 

Based on these processes, the star-formation history and metal-enrichment history of each galaxy are computed. We calculate the spectral 
energy distribution by synthesizing a simple stellar population model provided by \cite{2003MNRAS.344.1000B}. 
We calculate the dust extinction by assuming (1) the dust-to-gas ratio is proportional to the cold gas metallicity, (2) the optical depth of the dust is proportional 
to the column density of the dust, (3) the dust geometry follows the slab dust model \citep{1989MNRAS.239..939D}, and (4) the 
wavelength dependence of the attenuation obeys the Calzetti law \citep{2000ApJ...533..682C}. Our model also includes the evolution of 
the SMBHs and properties of active galactic nuclei (AGNs). 
See \cite{2019MNRAS.482.4846S} for details of the latest model of the SMBH growth and AGN properties.

\subsection{H$\alpha$ luminosity} \label{sec:LHa} 

We make a mock catalog of galaxies at $z=0.4$ in a (411.8~Mpc)$^3$ box based on the $\nu^2$ GC. The mock catalog 
includes 3,100,052 galaxies brighter than --15 mag in SDSS $r$-band. We calculate intrinsic luminosity of H$\alpha$ emission from 
each galaxy simply by converting from the SFR adopting 
the \cite{1998ARA&A..36..189K} relation corrected for the Chabrier IMF:
\begin{equation}
L_{{\rm H}\alpha}^{\rm int}~[{\rm erg~s}^{-1}] = SFR~[M_{\odot}~{\rm yr}^{-1}] / (4.4 \times 10^{-42}).
\end{equation}
Here we use the mean SFRs during dynamical times of disks and bulges for normal star-forming galaxies and starburst galaxies, 
respectively. 

The dust attenuation level of the nebular emission line compared to that of the continuum emission, $f_{\rm neb}$, is often quantified 
by the following expression:
\begin{equation}
f_{\rm neb} \equiv E(B-V)_{\rm star} / E(B-V)_{\rm line}
\end{equation}
\citep[e.g.,][]{2019PASJ...71....8K}. Various observational studies report that $f_{\rm neb}=0.44-0.6$ at $z<1$ 
\citep{2000ApJ...533..682C, 2011ApJ...738..106W, 2015ApJ...807..141P}.
In this study, we employ a fixed value of $f_{\rm neb}=0.5$. 
Including the dust extinction, the observable luminosity of H$\alpha$ emission, $L_{{\rm H}\alpha}^{\rm obs}$, is calculated.
Here we note that some previous observational studies at higher redshift ($z>0.8$) showed that AGN-powered HAEs could 
contribute to the luminous HAE sample \citep[e.g.,][]{2016MNRAS.457.1739S, 2017MNRAS.471..629M}. 
According to \cite{2016MNRAS.457.1739S}, the number fraction of AGNs in luminous HAEs at $z=0.8-2.2$ is $\sim$30\%.
They also found that the AGN fraction increases with $L_{\rm H\alpha}$ at $L_{\rm H\alpha} \geq 10^{42}$~erg~s$^{-1}$.
This trend is constant independent of the redshifts (at least at $z=0.8-2.3$). 
Although, at high redshifts, the mechanism of AGN that enhances the Ha luminosity may play an important role on the H$\alpha$ 
LF at the bright end, at $z\sim0.4$, the HAEs at $L_{\rm H\alpha} \geq 10^{42}$~erg~s$^{-1}$ are relatively rare
\citep[e.g., ][]{2009A&A...507..781S, 2018PASJ...70S..17H}.
 
Since we here focus on the field variance of the LFs of HAEs much fainter than the bright end, we do not include the contribution 
from the HAEs powered by the AGNs.

\subsection{HAE sample} \label{sec:HAE} 

In this subsection, we describe the definition of HAEs in our model. 
We define the HAE sample by adopting a cut by the rest-frame EW ($EW_0$) of the H$\alpha$ emission line.
The $EW_0$ is defined as follows:
\begin{equation}
EW_0 \equiv \frac{L_{\rm H\alpha}}{L_{\lambda6563}},
\end{equation}
where $L_{\rm H\alpha}$ and $L_{\lambda6563}$ are the integrated luminosity of H$\alpha$ emission and the
continuum luminosity at the wavelength of the H$\alpha$ line.
We employ $EW_0 \geq$ 40~{\AA} which is the same value as that adopted in \cite{2018PASJ...70S..17H}.
The H$\alpha$ LFs with other EW cuts (see Figure \ref{fig:LF_variousEW}) are described in Section \ref{sec:HaLF}.

It should be noted that, in the $\nu^2$GC, almost all the gas in a starburst galaxy is consumed at the end of the burst.
Moreover, the H$\alpha$ emission line is radiated from the ionized gas in the H~{\sc ii} region of galaxies. Therefore, 
when a starburst galaxy do not have sufficient gas, H$\alpha$ emission is not radiated. 
To eliminate such a gas-poor galaxy from the HAE sample, we employ further cut by the cold gas fraction, $f_{\rm gas}$, 
of starburst galaxies, defined by the following expression:
\begin{equation}
f_{\rm gas} \equiv \frac{ M_{\rm gas} }{ M_{\rm gas} + M_{\rm star} },
\end{equation}
where $M_{\rm gas}$ and $M_{\rm star}$ are  cold gas mass and stellar mass, respectively. We set $f_{\rm gas}$ to 
reproduce observed H$\alpha$ LFs at $z=0.4$, $f_{\rm gas} \geq 0.10$. With the current observational dataset, it is 
difficult to determine the precise value of the $f_{\rm gas}$ cut. The $f_{\rm gas}$ cut affects the shape of the bright 
end of the H$\alpha$ LF (Figure \ref{fig:LF_various_fgas}). In this study, we specifically focus on the field variance on 
the H$\alpha$ LF and the shape of the bright end does not have an impact on the discussion. 
Therefore, $f_{\rm gas} \geq 0.10$ is a tentative threshold. The dependence of the cold gas fraction $f_{\rm gas}$ on 
the H$\alpha$ LF is discussed in Section \ref{sec:HaLF}. Based on these  considerations, we obtain model HAE sample 
consisting of 407,353 HAEs with $L_{\rm H\alpha} \geq 10^{40}$~erg~s$^{-1}$.
We describe physical properties of the $\nu^2$GC HAEs in Section \ref{sec:HAEprop}.

\section{Physical properties of $\nu^2$GC HAEs} \label{sec:HAEprop} 

\subsection{The H$\alpha$ LF} \label{sec:HaLF} 

\begin{figure*} 
\begin{center}
\includegraphics[width=88mm]{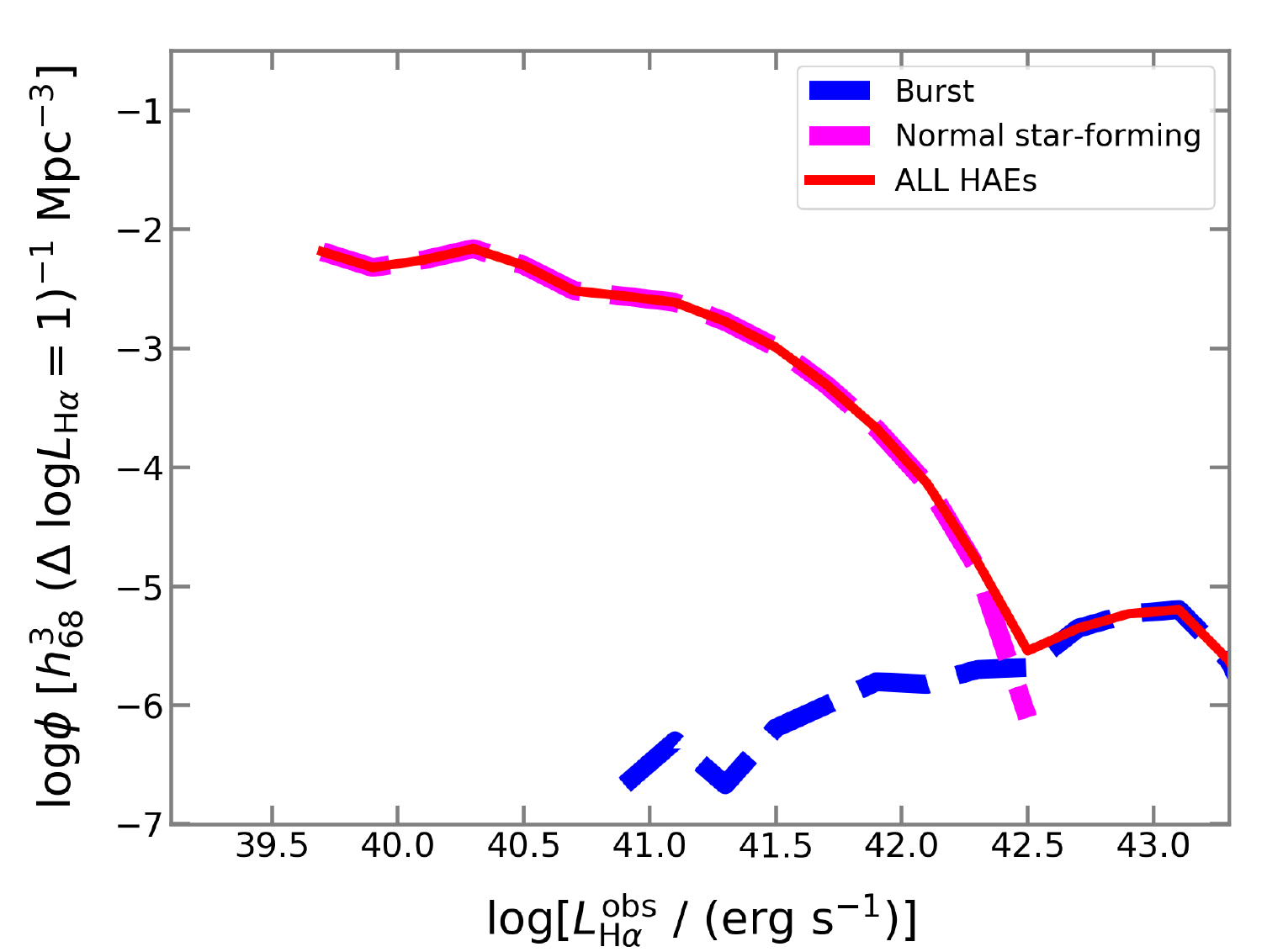}
\includegraphics[width=88mm]{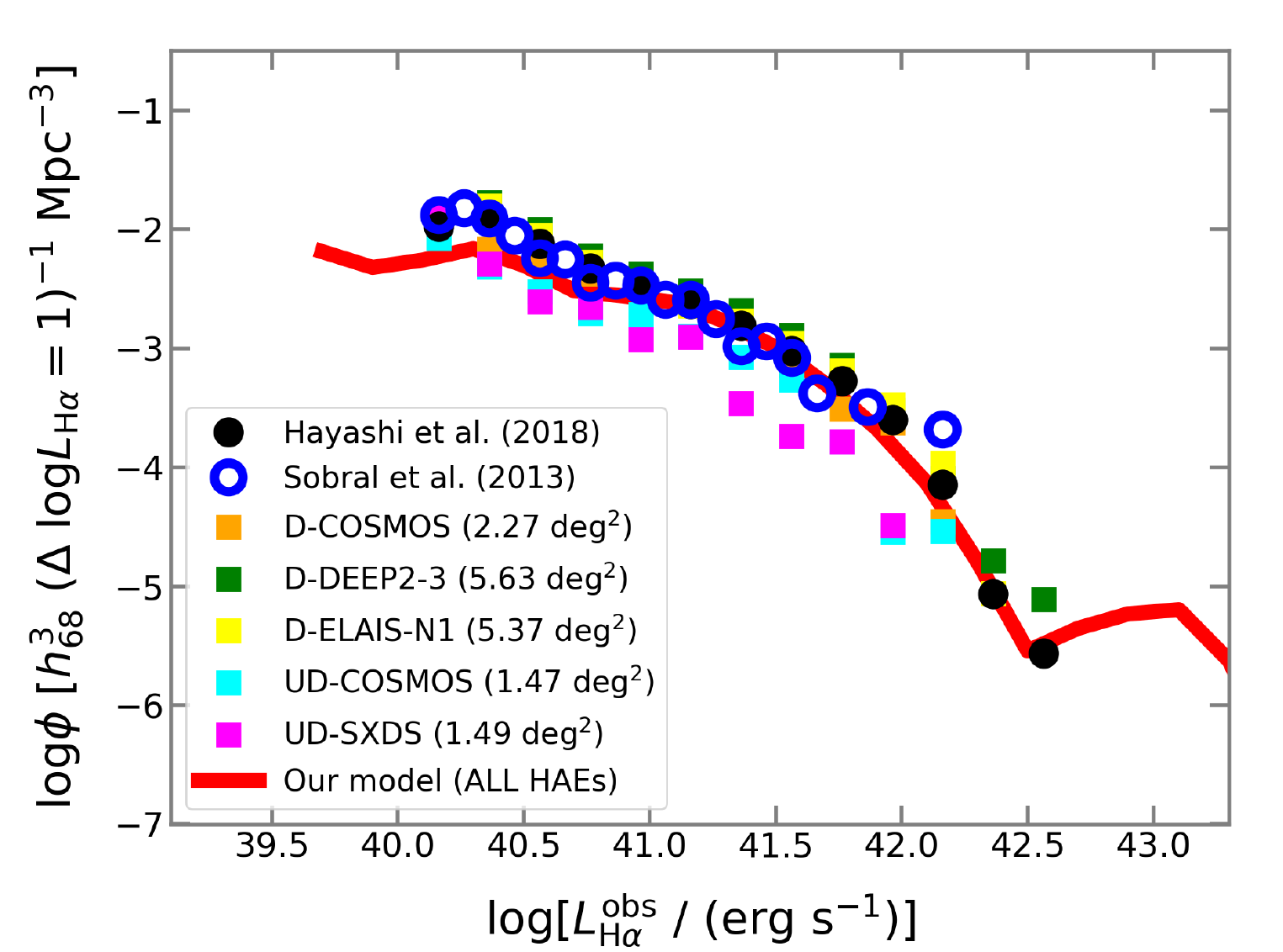}
\caption{
The H$\alpha$ LF (with the dust extinction) of the model HAEs (red solid line). {\it Left}: The H$\alpha$ LF of starburst HAEs 
(blue dashed line) and normal star-forming HAEs (magenta dashed line) are superimporsed. 
{\it Right}: Comparing the model H$\alpha$ LF with observed ones (non-dust corrected) at $z=0.4$ obtained by HiZELS 
\citep[][blue open circles]{2013MNRAS.428.1128S} and HSC-SSP \citep[][black filled circles: all the survey fields, colored squares: 
individual survey field]{2018PASJ...70S..17H}.
Note that \cite{2013MNRAS.428.1128S} corrected dust attenuation by a simple way assuming 1 mag extinction for all HAEs. 
Thus we have plotted the observed LF by offsetting the corrected LF in \cite{2013MNRAS.428.1128S}; i.e., we have subtracted 
0.4 dex from log$L_{\rm H\alpha}$ in their Table 4.
\label{fig:LF_basic} }
\end{center}
\end{figure*}

\begin{figure} 
\begin{center}
\includegraphics[width=88mm]{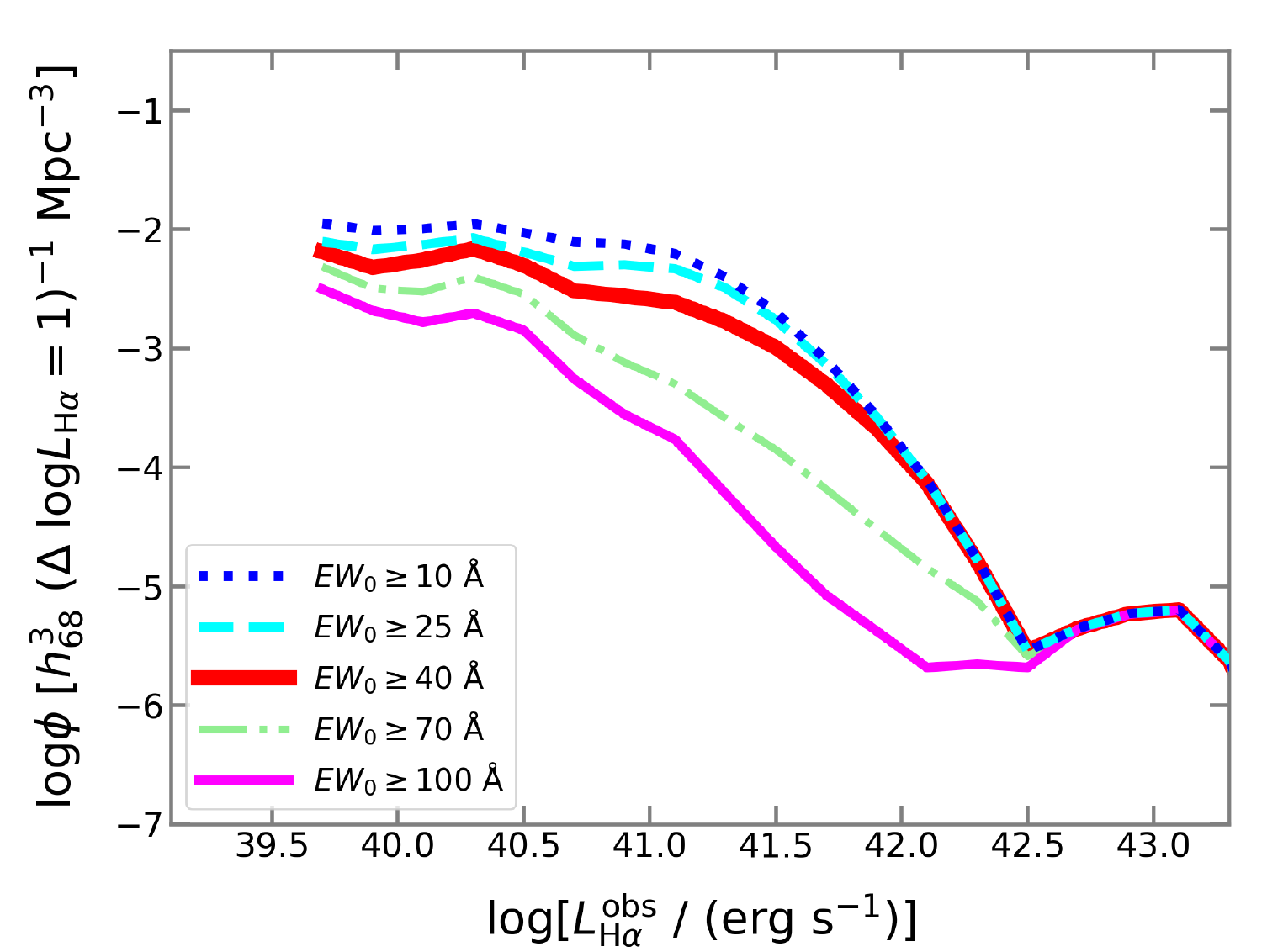}
\caption{
The H$\alpha$ LFs with various EW cuts. The red thick solid line show our standard model
with $EW_0 \geq 40$ {\AA}. Thin lines indicate the LFs when we adopt other EW cut; specifically,  
blue dotted: $EW_0 \geq 10$ {\AA} \citep[similar value to the observational study by][]{2007ApJ...657..738L},
cyan dashed: $EW_0 \geq 25$ {\AA} \citep[the same as][]{2013MNRAS.428.1128S},
green dashed-dotted: $EW_0 \geq 70$ {\AA},
and magenta solid: $EW_0 \geq 100$ {\AA} \citep[the same as][]{2013MNRAS.433..796D}. 
Other parameters are fixed; $f_{\rm neb}=0.5$ and $f_{\rm gas} \geq 0.10$.
\label{fig:LF_variousEW} }
\end{center}
\end{figure}

\begin{figure} 
\begin{center}
\includegraphics[width=88mm]{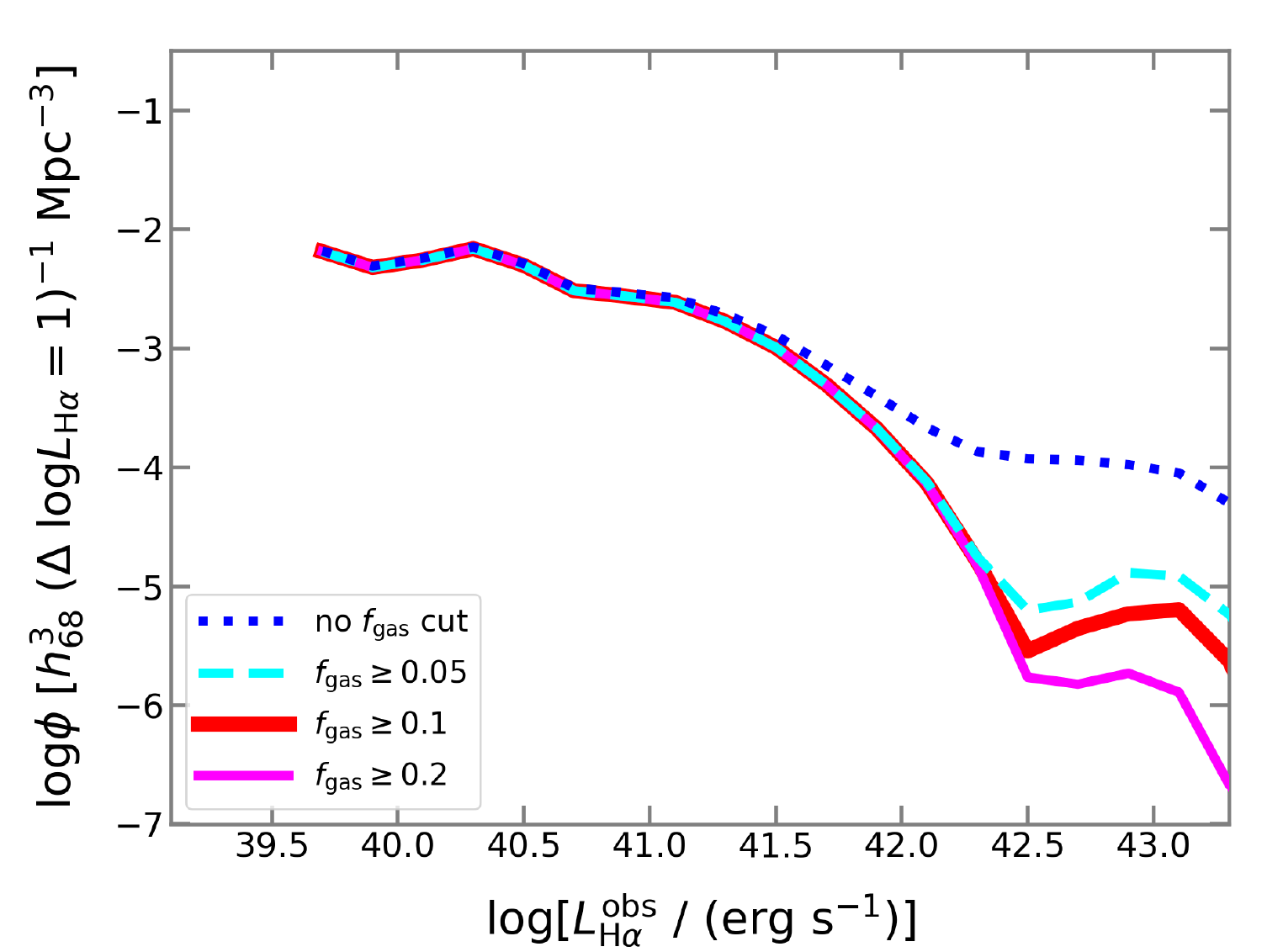}
\caption{
The H$\alpha$ LFs with various $f_{\rm gas}$ cuts. 
The thick solid red line show our standard model with $f_{\rm gas} \geq 0.10$. 
Thin lines indicate the LFs when we adopt other $f_{\rm gas}$ cut; specifically,  
blue dotted: without $f_{\rm gas}$ cut,
cyan dashed: $f_{\rm gas} \geq 0.05$,
and magenta solid: $f_{\rm gas} \geq 0.20$. 
Other parameters are fixed; $f_{\rm neb}=0.5$ and $EW_0 \geq 40$ {\AA}.
\label{fig:LF_various_fgas} }
\end{center}
\end{figure}

\begin{figure} 
\begin{center}
\includegraphics[width=88mm]{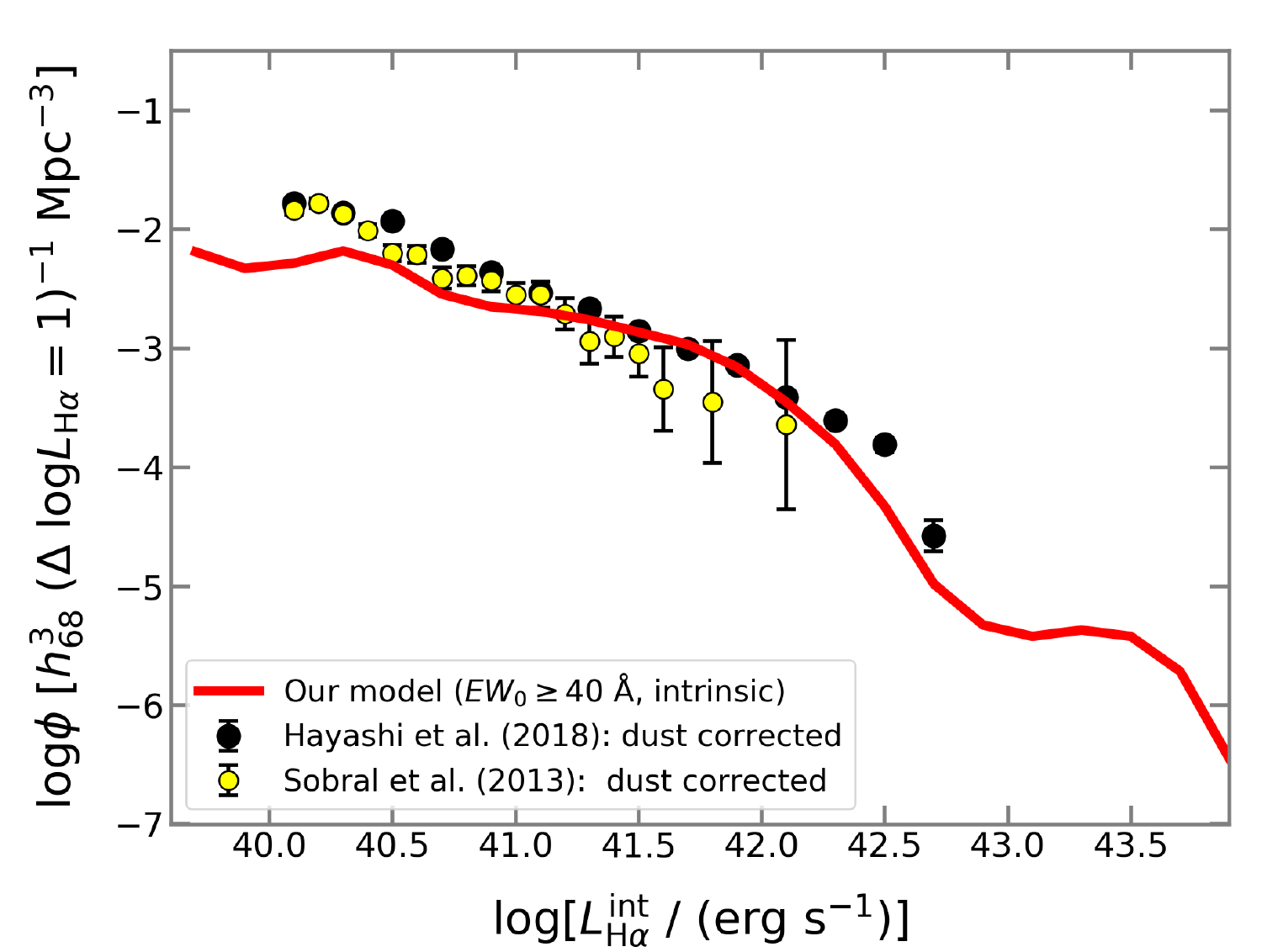}
\caption{
The intrinsic H$\alpha$ LF of the model HAEs (red solid line). Black and yellow filled circles show the 
observed LF corrected for the dust attenuation obtained by the HSC-SSP \citep[][]{2018PASJ...70S..17H}
and HiZELS \citep[][]{2013MNRAS.428.1128S}, respectively.
\label{fig:LF_int} }
\end{center}
\end{figure}

The left panel of Figure \ref{fig:LF_basic} shows the H$\alpha$ LF derived from our HAE model. 
The red solid line shows the H$\alpha$ LF derived by using all HAEs while magenta and blue dashed lines show H$\alpha$ LFs of 
normal star-forming HAEs and  HAEs during starburst (hereafter starburst HAEs). As seen in the figure, normal star-forming HAEs 
dominate the fainter regime ($L_{\rm H\alpha} \lesssim 10^{42.5}$~erg~s$^{-1}$) of the H$\alpha$ LF while starburst HAEs contribute 
the bright end of the LF.

In the right panel of Figure \ref{fig:LF_basic}, we compare model H$\alpha$ LF (red solid line) with observed ones.
Blue open circles show H$\alpha$ LF obtained by High-redshift(Z) Emission Line Survey \citep[HiZELS;][]{2013MNRAS.428.1128S}. 
The survey area of \cite{2013MNRAS.428.1128S} is 2~deg$^2$. 
The filled circles and colored squares show H$\alpha$ LFs obtained by the HSC-SSP \citep[][]{2018PASJ...70S..17H}.
The HSC-SSP HAE sample is obtained by utilizing NB921 images taken in two UltraDeep (UD) fields (UD-COSMOS and UD-SXDS
fields) and three Deep (D) fields (D-COSMOS, D-DEEP2-3, and E-ELAIS-N1 fields). See \citet{2018PASJ...70S...8A} and 
\citet{2018PASJ...70S..17H} for further details of the survey. The H$\alpha$ LF shown by black filled circles is derived by using all HAEs 
in all of the survey fields (16.2 deg$^2$) and those shown by colored squares are achieved in each individual field.

As seen in the figure, {\it the observed H$\alpha$ LF in the individual HSC fields show a scatter up to $\sim$1~dex.}
The H$\alpha$ LFs from \cite{2013MNRAS.428.1128S} and all of the survey fields of \cite{2018PASJ...70S..17H} show no significant 
differences. Surprisingly, these two observational studies adopt different H$\alpha$ $EW_0$ thresholds to select HAEs, 
$EW_0 \geq 40$~{\AA} in \cite{2018PASJ...70S..17H} and $EW_0 \geq 25$~{\AA} in \cite{2013MNRAS.428.1128S}.
Figure \ref{fig:LF_variousEW} shows the model H$\alpha$ LFs with various EW cuts. When we adopt smaller EW cuts [10~{\AA} and 25~{\AA}; 
similar values to \cite{2007ApJ...657..738L} and \cite{2013MNRAS.428.1128S}, respectively], the H$\alpha$ LF does not show significant 
difference (up to $\sim$0.3~dex depending on the luminosity range; similar to the field variance as shown in Section \ref{sec:discussion}), 
while the number density of HAEs is significantly lower than observed one 
in the case that we adopt $EW_0 \geq$ 100~{\AA} \citep[the same as][]{2013MNRAS.433..796D}.
{\it Presumably, the reason for these scatters and apparent agreement are given by the field variance.} 
We discuss how the field variance affects the H$\alpha$ LF in Section \ref{sec:discussion}.

In Figure \ref{fig:LF_various_fgas}, we show the dependence of the $f_{\rm gas}$ cut on the H$\alpha$ LF.
Without the $f_{\rm gas}$ cut, the number density of model HAEs with $L_{\rm H\alpha} \gtrsim 10^{42}$~erg~s$^{-1}$ is significantly
higher than observed one. When we apply $f_{\rm gas} \geq$ 0.05 -- 0.20, the H$\alpha$ LF shows no significant differences at  
$L_{\rm H\alpha} \lesssim 10^{42.3}$~erg~s$^{-1}$. Based on the current observational dataset, we cannot restrict the bright end
of the H$\alpha$ LF. When we discuss the field variance on the H$\alpha$ LF in Section \ref{sec:discussion}, we focus on the number 
density of HAEs at $L_{\rm H\alpha} = 10^{40.5}$ and $10^{41.5}$~erg~s$^{-1}$ in which the LF is independent of the $f_{\rm gas}$ cut.

Figure \ref{fig:LF_int} shows the intrinsic H$\alpha$ LF in the model compared with the dust corrected LF obtained by the HSC-SSP 
\citep{2018PASJ...70S..17H} and HiZELS \citep[][]{2013MNRAS.428.1128S}. Here model HAEs are selected by dust attenuated 
H$\alpha$ $EW_0$ and the intrinsic LF is derived by their intrinsic H$\alpha$ luminosities. 
Our model shows general agreement with the dust corrected LF of observed HAEs while the number density of model HAEs is 
slightly lower than those of observational results at the faint regime ($L_{\rm H\alpha} \lesssim 10^{41}$~erg~s$^{-1}$).
This difference may be due to difference in the extinction curves. 
As we described in subsection \ref{sec:nu2gc}, the Calzetti law \citep{2000ApJ...533..682C} is used in the $\nu^2$GC.
On the other hand, \cite{2018PASJ...70S..17H} estimate the amount of dust extinction based on the Balmer decrement
by assuming the extinction curve by \cite{1989ApJ...345..245C}. Moreover, \cite{2013MNRAS.428.1128S} corrected dust attenuation by a 
simple way assuming 1 mag extinction for all HAEs. 
Since these differences have no significant impacts on the conclusion on the field variance, more detailed discussions will be a future work.

\subsection{Model HAEs in the log($SFR$) -- log($M_{\rm star}$) plane} \label{sec:MS} 

We examine the relation between the SFR and stellar mass of model HAEs. Various previous observations have reported that 
star-forming galaxies show a tight correlation between their SFR and stellar mass, which is called the ``main sequence (MS)'' 
of star-forming galaxies \citep[e.g.,][]{2007A&A...468...33E, 2007ApJ...660L..43N, 2009MNRAS.394....3D, 2009A&A...504..751S, 
2013ApJ...777L...8K, 2014ApJS..214...15S, 2018A&A...619A..27B}. Therefore it is a good test to show the distribution 
of galaxies in the log(SFR) -- log($M_{\rm star}$) plane for the reliability of our model.
 
\begin{figure*} 
\begin{center}
\includegraphics[width=88mm]{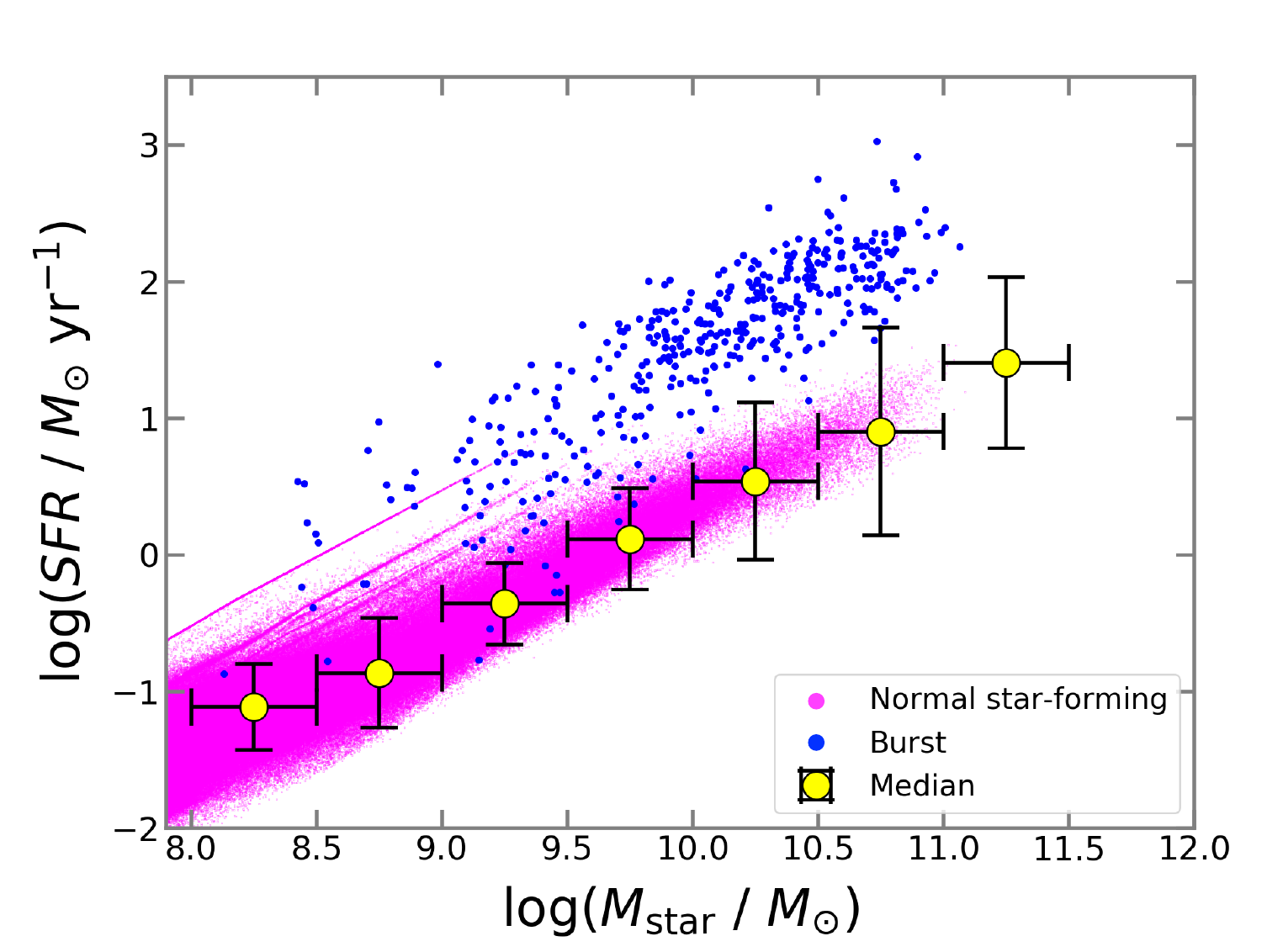}
\includegraphics[width=88mm]{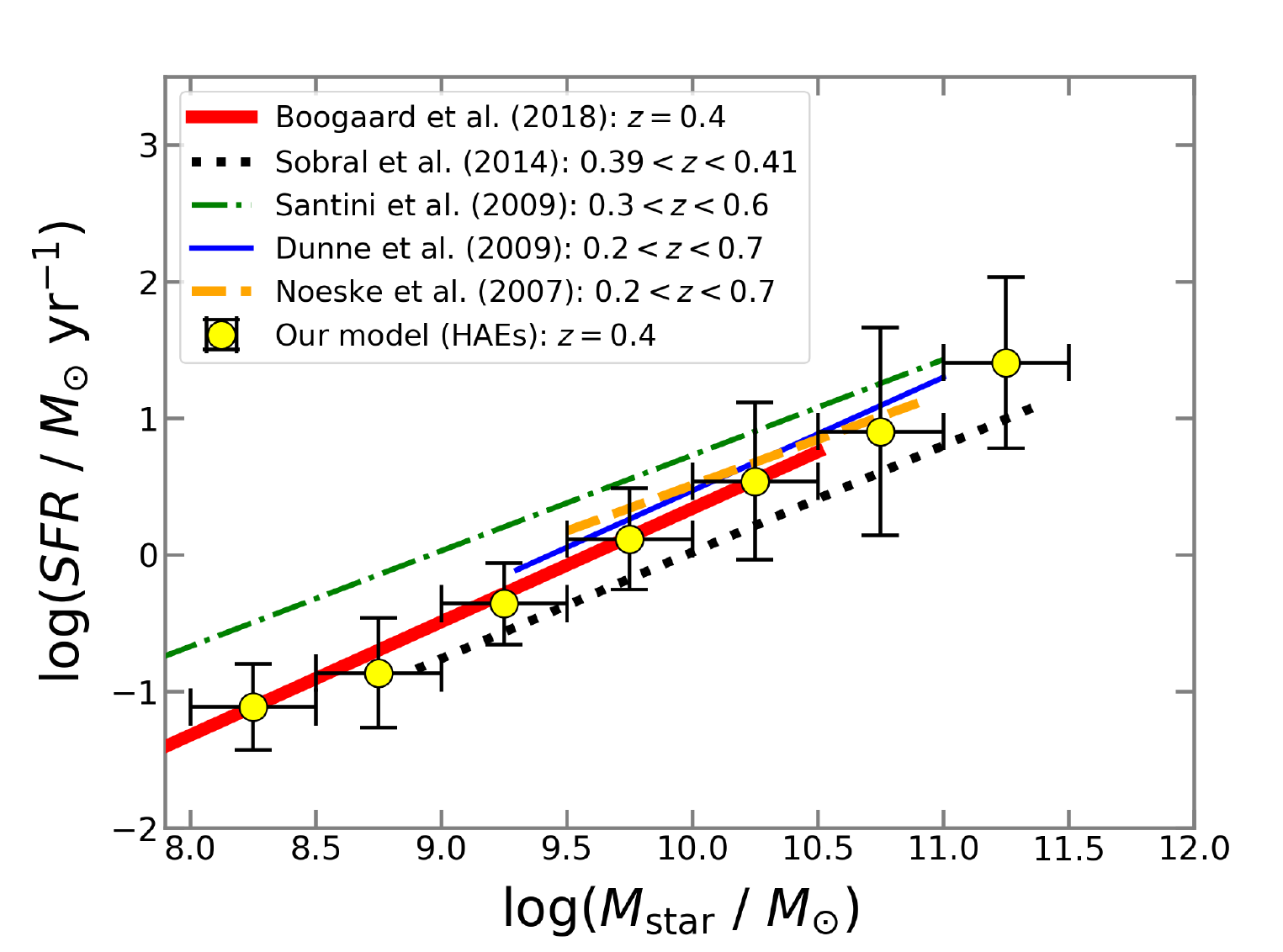}
\caption{
The log(SFR) -- log($M_{\rm star}$) plane of HAEs($EW_0 \geq 40$ {\AA}).
{\it Left}: Blue and magenta dots show starburst galaxies and normal star-forming galaxies, respectively. 
Yellow filled circles show median SFR in $M_{\rm star}$ bin ($\Delta$log$M_{\rm star}=0.5$). 
Error bars correspond to 1$\sigma$ standard deviation.
Right: Comparison of SFR versus stellar mass relation of HAEs between our model and observational results. 
Observational results are shown by colored lines; 
red thick solid: \cite{2018A&A...619A..27B} ($z = 0.4$), black dotted: \cite{2014MNRAS.437.3516S} ($0.39 < z < 0.41$),
green dashed-and-dotted: \cite{2009A&A...504..751S} ($0.3 < z < 0.6$),
blue thin solid: \cite{2009MNRAS.394....3D} ($0.2 < z < 0.7$), 
orange dashed: \cite{2007ApJ...660L..43N} ($0.2 < z < 0.7$).
\label{fig:MS} }
\end{center}
\end{figure*}

The left panel of Figure \ref{fig:MS} shows the log($SFR$) -- log($M_{\rm star}$) plane of model HAEs. 
While normal star-forming HAEs are shown by magenta dots, starburst HAEs are denoted by blue circles.
Model HAEs clearly show a tight correlation between SFR and stellar mass. 
The relation between the SFR and stellar mass can be fitted by an analytic function, 
\begin{equation}
{\rm log}[SFR / M_{\odot} {\rm yr}^{-1}] = \alpha_{\rm MS} {\rm log}[M_{\rm star} / M_{\odot}] + \beta_{\rm MS}. 
\end{equation}
The best-fit parameters are $\alpha_{\rm MS}=0.86\pm0.02$ and $\beta_{\rm MS}=-8.26 \pm 0.24$.
The scatter in the log(SFR) direction is $\sigma_{\rm MS}=0.48\pm0.18$.
We compare our model HAE to observed MS of star-forming galaxies at similar redshift in the log($SFR$) -- log($M_{\rm star}$) 
plane as shown in the right panel of Figure \ref{fig:MS}. The model well reproduces the MS of star-forming galaxies
[\cite{2018A&A...619A..27B} at $z=0.4$: $\alpha_{\rm MS}=0.83$, $\beta_{\rm MS}=-7.96$ and $\sigma_{\rm MS}=0.44$;
\cite{2014MNRAS.437.3516S} at $0.39<z<0.41$: $\alpha_{\rm MS}=0.78$, $\beta_{\rm MS}=-7.78$ and $\sigma_{\rm MS}=0.49$].
Since model HAEs distribute along the observed main sequence, it is suggested that HAEs are typical MS galaxies.

\subsection{The spatial distribution} \label{subsec:spatial} 

\begin{figure*} 
\begin{center}
\includegraphics[width=185mm]{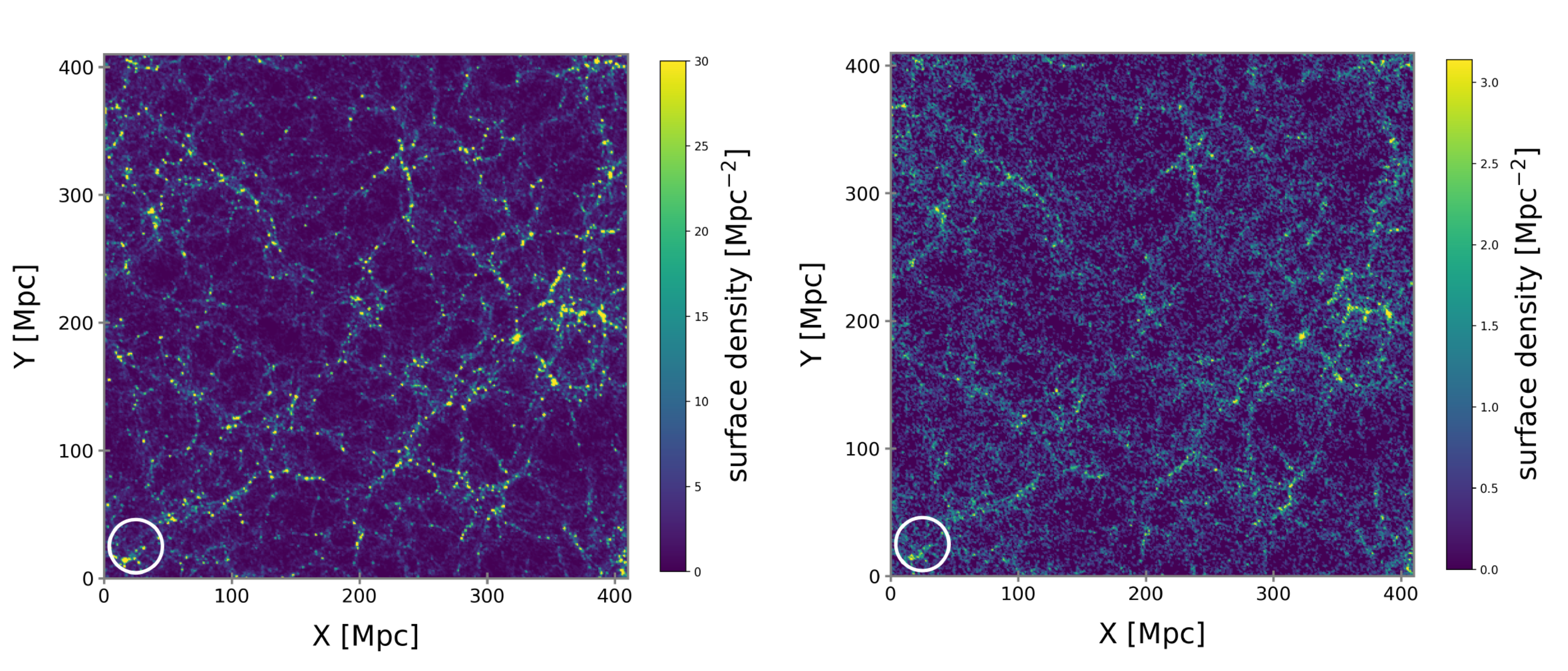}
\caption{The surface number density distribution map of galaxies with $r<-15$ (left) and HAEs with 
$L_{\rm H\alpha} \geq 10^{40}$~erg~s$^{-1}$ (right). White circles at left bottom in each map
show the field-of-view of Subaru/HSC (1.5 degree diameter).
\label{fig:densitymap} }
\end{center}
\end{figure*}

We investigate how the HAE selection affects the spatial distribution of galaxies. Figure \ref{fig:densitymap} 
shows the surface number density distribution of model galaxies with $r<-15$ mag (left) and HAEs with 
$L_{\rm H\alpha} \geq 10^{40}$~erg~s$^{-1}$ (right) over $411.8 \times 411.8$~Mpc$^2$ (corresponding to
14.8 $\times$ 14.8~deg$^2$). We project a 411.8~Mpc $\times$ 411.8~Mpc region with a thickness of 70~Mpc 
(corresponding to the width of the HSC NB921 filter which is used to observe HAEs at $z=0.4$) to 
make a 2-dimensional image. Within a 411.8~Mpc $\times$ 411.8~Mpc $\times$ 70~Mpc region, there are  
537,190 galaxies (SDSS $r<-15$ mag) and 67,808 HAEs ($L_{\rm H\alpha} \geq 10^{40}$~erg~s$^{-1}$). 
The average surface number density of HAEs with $L_{\rm H\alpha} \geq 10^{40}$~erg~s$^{-1}$ within a 
411.8~Mpc $\times$ 411.8~Mpc $\times$ 70~Mpc box is 308.9 deg$^{-2}$.
The local surface number density distribution is calculated by counting the number of galaxies within a fixed 
aperture with a radius of 1 Mpc. Both of all galaxies and HAEs show clear filamentary structure.

\begin{figure*} 
\begin{center}
\includegraphics[width=185mm]{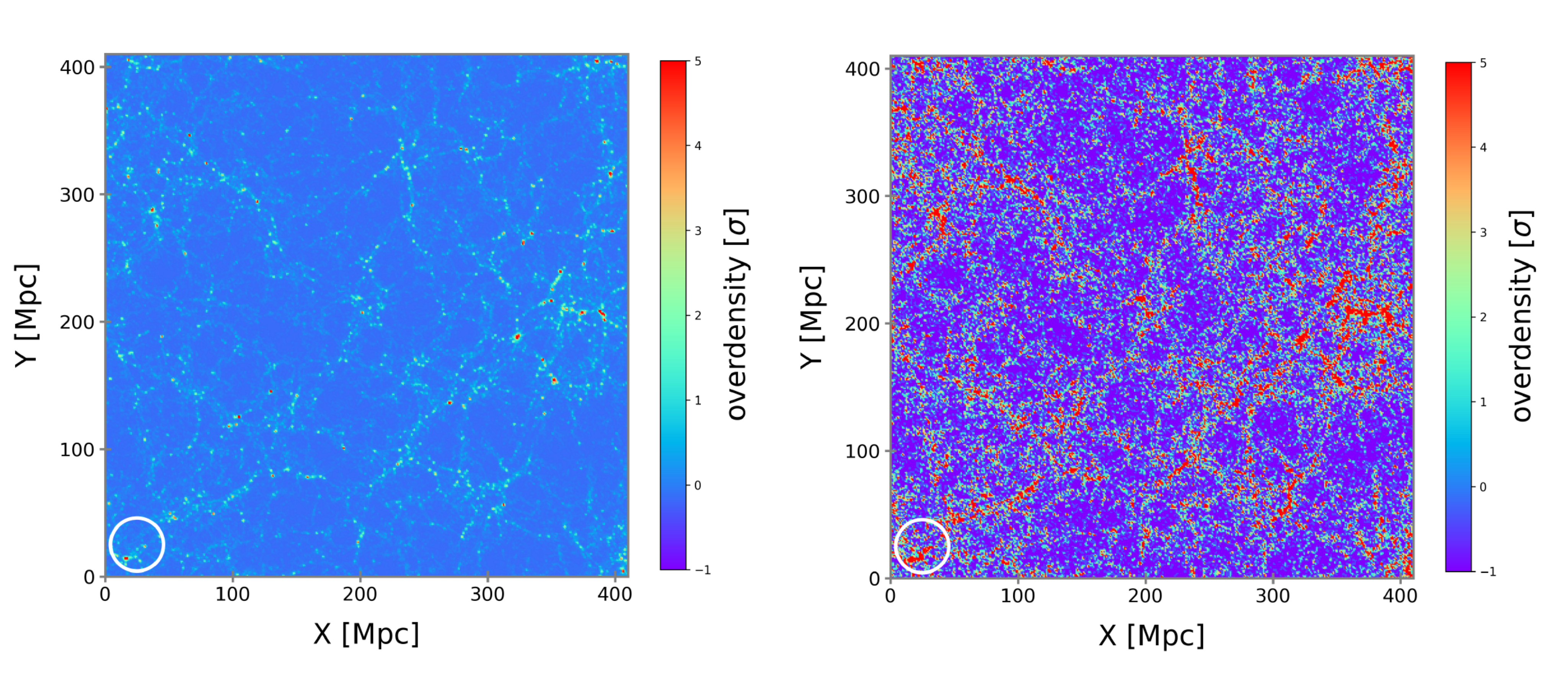}
\caption{Overdensity ($\delta$) map of galaxies with $r<-15$ (left) and HAEs (right) with 
$L_{\rm H\alpha} \geq 10^{40}$~erg~s$^{-1}$. Regions where the $\delta$ exceeds 5$\sigma$
are shown by red color in each map. White circles at left bottom in each map show the 
field-of-view of Subaru/HSC (1.5 degree diameter).
\label{fig:sigmamap} }
\end{center}
\end{figure*}

We then examine the overdensity $\delta$ distribution to investigate how HAEs trace the structure.
The overdensity $\delta$ is defined as follows:
\begin{equation}
\delta \equiv \frac{ n - \bar{n} }{ \bar{n} },
\end{equation}
where $n$ is the surface number density of galaxies within an aperture and $\bar{n}$ is the mean surface
number density over the whole projected area.
In Figure \ref{fig:sigmamap} we show the $\delta$ map of galaxies with $r<-15$ mag (left) and HAEs with
$L_{\rm H\alpha} \geq 10^{40}$~erg~s$^{-1}$ (right). The average and median of the surface number density of 
galaxies in an aperture are 3.04~Mpc$^{-2}$ and 1.59~Mpc$^{-2}$, while those of HAEs are 0.40~Mpc$^{-2}$ and
0.32~Mpc$^{-2}$.
In each panel of Figure \ref{fig:sigmamap}, regions in which the overdensity exceeds 5$\sigma$
are colored red. Clearly, the spatial distribution of HAEs shows higher overdensity in galaxy clusters and 
the filamentary structures compared to that of all galaxies, suggesting that HAEs are a good tracer to 
investigate structures such as cosmic filaments.
We will examine the spatial distribution in more detail, including the clustering analysis, in a future paper.

\section{Field variance on the H$\alpha$ LF} \label{sec:discussion} 

\begin{figure*} 
\begin{center}
\includegraphics[width=160mm]{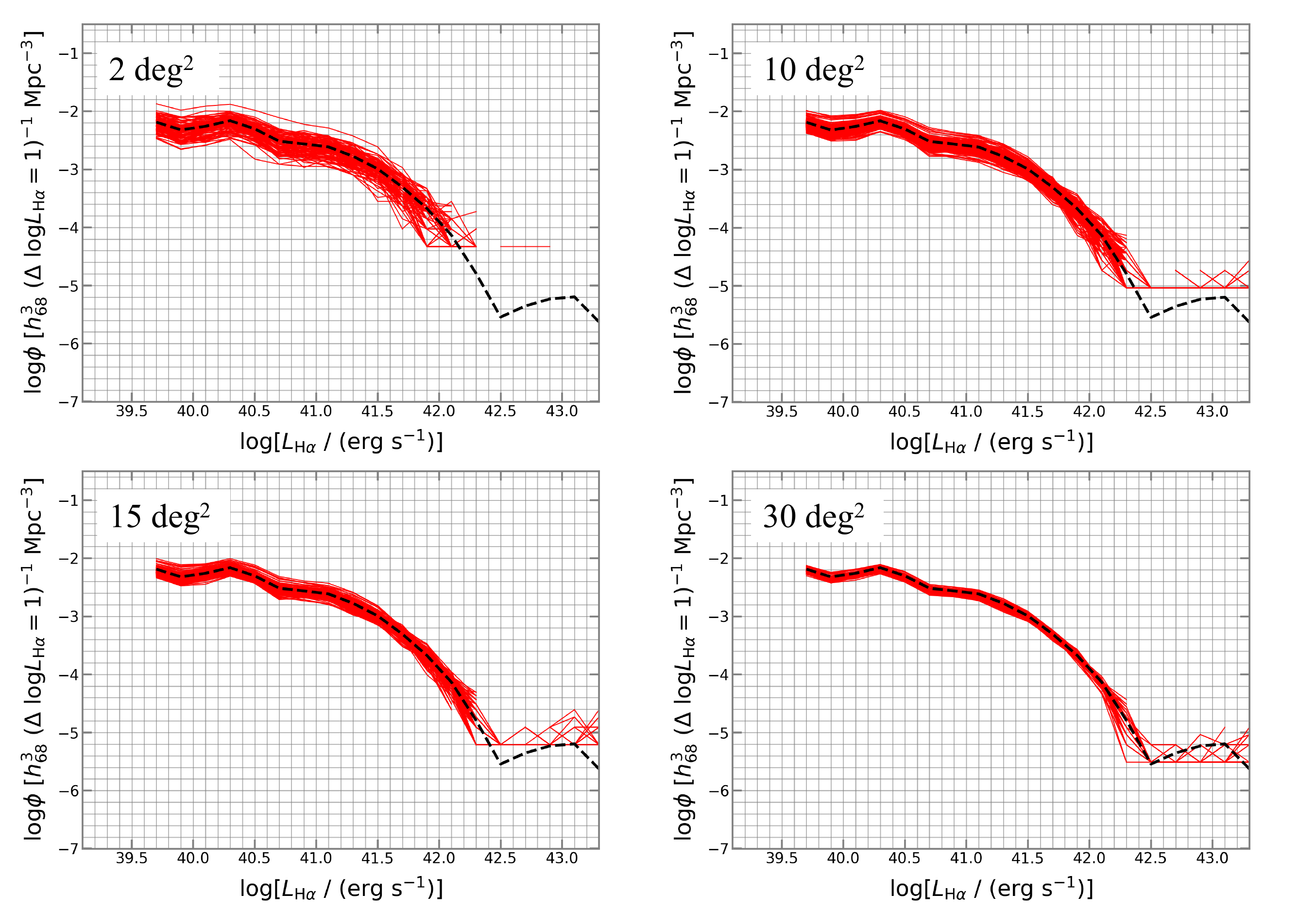}
\caption{H$\alpha$ LFs within various survey area (red lines). Black dashed line in each panel
shows the LF derived based on all HAEs in the (411.8 Mpc)$^3$ box.
2, 10, 15, and 30~deg$^2$ correspond to $\sim$1,521, 3,844, 11,556, and 23,165~Mpc$^2$ in the
comoving scale at $z=0.4$. 
\label{fig:LF_env} }
\end{center}
\end{figure*}

\begin{figure*} 
\begin{center}
\includegraphics[width=88mm]{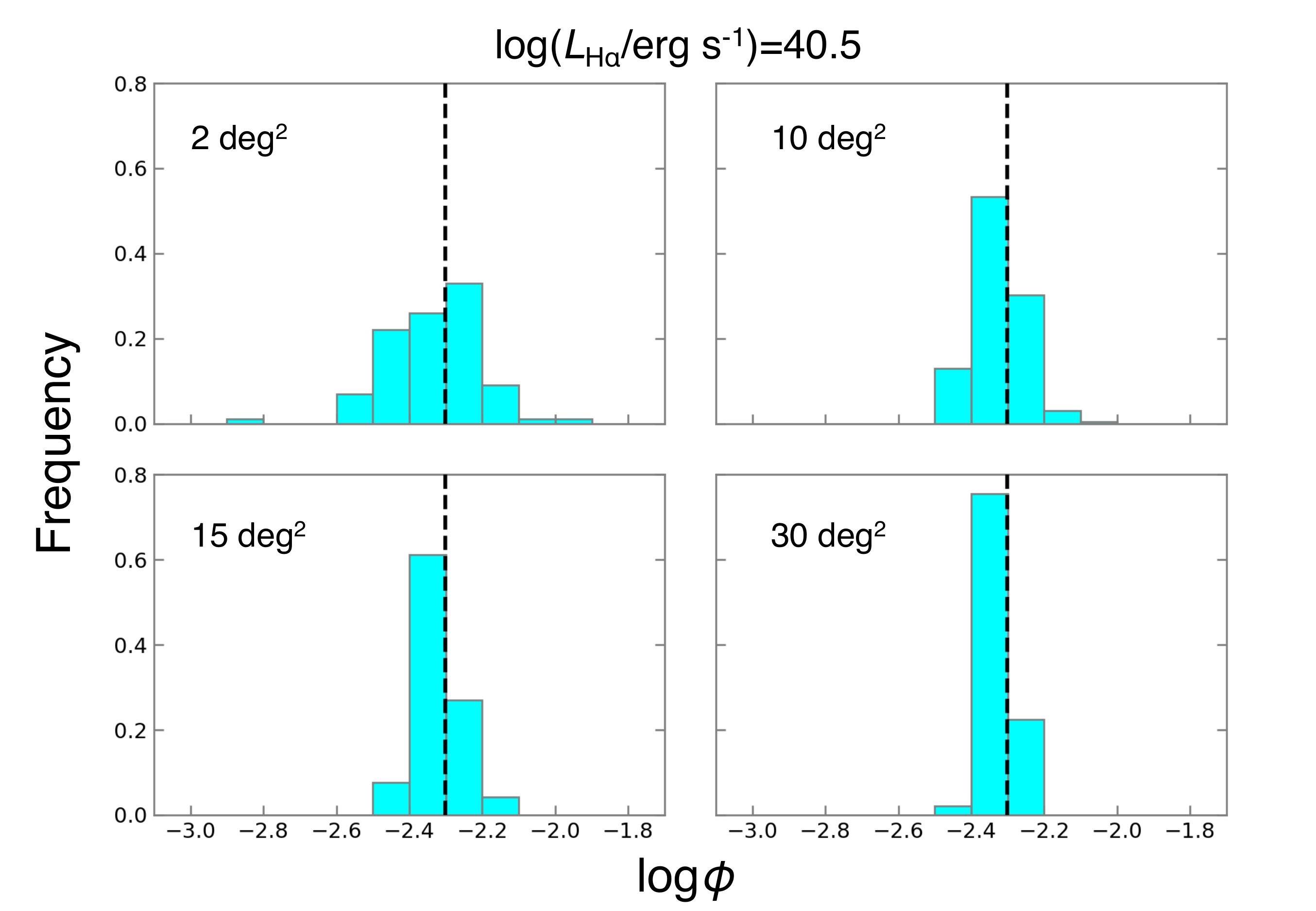}
\includegraphics[width=88mm]{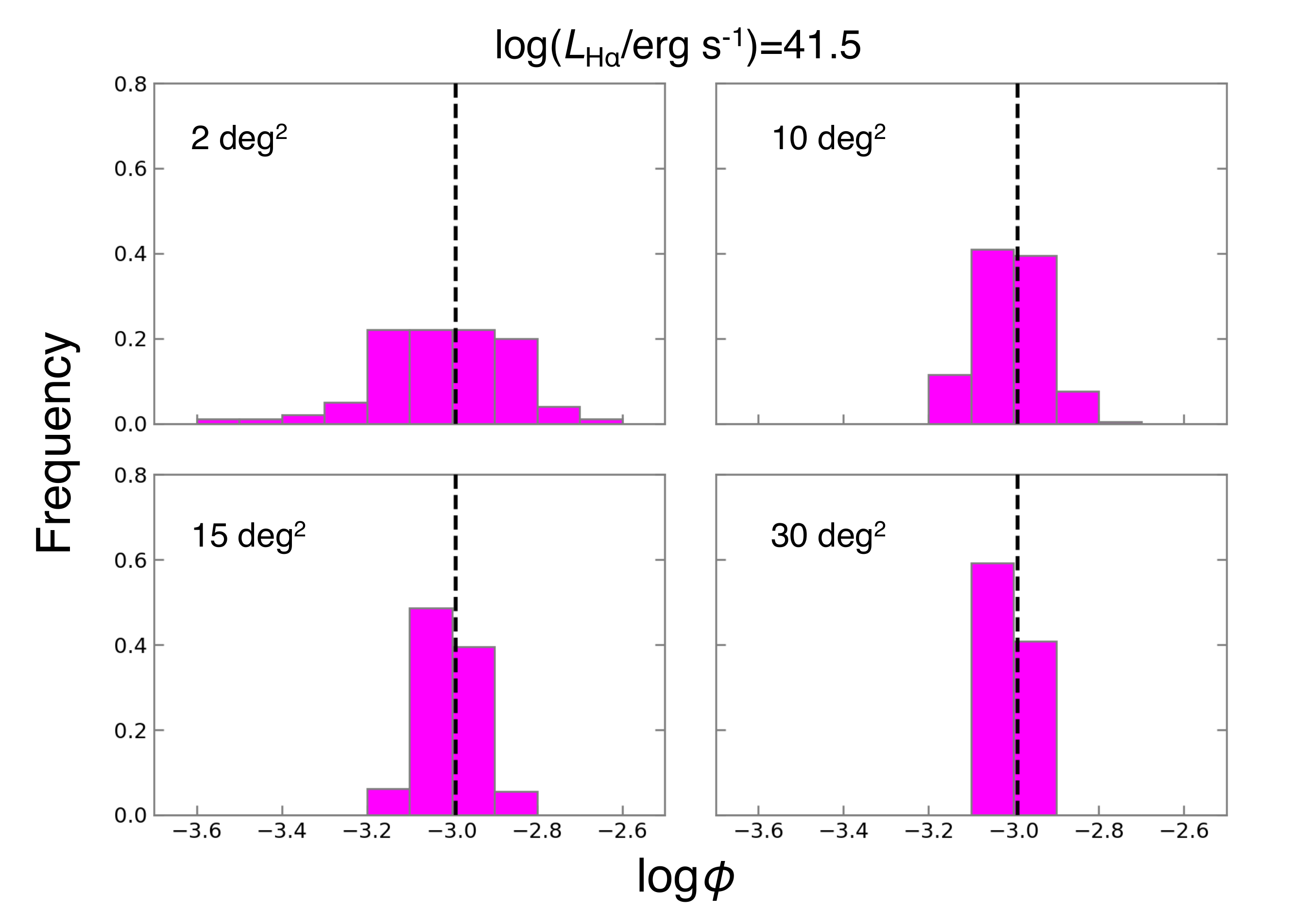}
\caption{
The frequency distribution of $\phi$ at $L_{\rm H\alpha}=10^{40.5}$~erg~s$^{-1}$ (left panels)
and $L_{\rm H\alpha}=10^{41.5}$~erg~s$^{-1}$ (right panels).  The survey areas shown here are the
same as Figure \ref{fig:LF_env}. Vertical dashed line in each panel indicate the number density of 
HAEs in whole of (411.8~Mpc)$^3$ box. 
\label{fig:phi_hist} }
\end{center}
\end{figure*}

In this section we discuss the field variance of the HAE distribution by focusing on the fluctuation of H$\alpha$ LFs in 
various environments. For this purpose, we divide the 411.8~Mpc $\times$ 411.8~Mpc field into square subregions and 
derive H$\alpha$ LFs in those regions. Figure \ref{fig:LF_env} shows H$\alpha$ LFs derived by applying various areas 
of square region. 
In Figure \ref{fig:phi_hist}, we show the frequency distribution of HAE number density, $\phi$, at $L_{\rm H\alpha} = 
10^{40.5}$~erg~s$^{-1}$ (left) and at $10^{41.5}$~erg~s$^{-1}$ (right). 
The spread of the LFs decreases with increasing survey the areas (see also Table \ref{tab:LF_MinMaxStdev}).
In the case of 2~deg$^2$ region (typical survey area of previous observations), the H$\alpha$ LF 
shows significantly large variance 
up to $\sim$1~dex. Therefore, any surveys narrower than 2~deg$^2$ could contain uncertainty of the number density 
at least $\sim$1~dex caused by the field variance.
Furthermore, if a rare bright HAE is observed, the bright end of the LF could be overestimated by at least one order of 
magnitude, due to a small survey area. 

\begin{figure*} 
\begin{center}
\includegraphics[width=85mm]{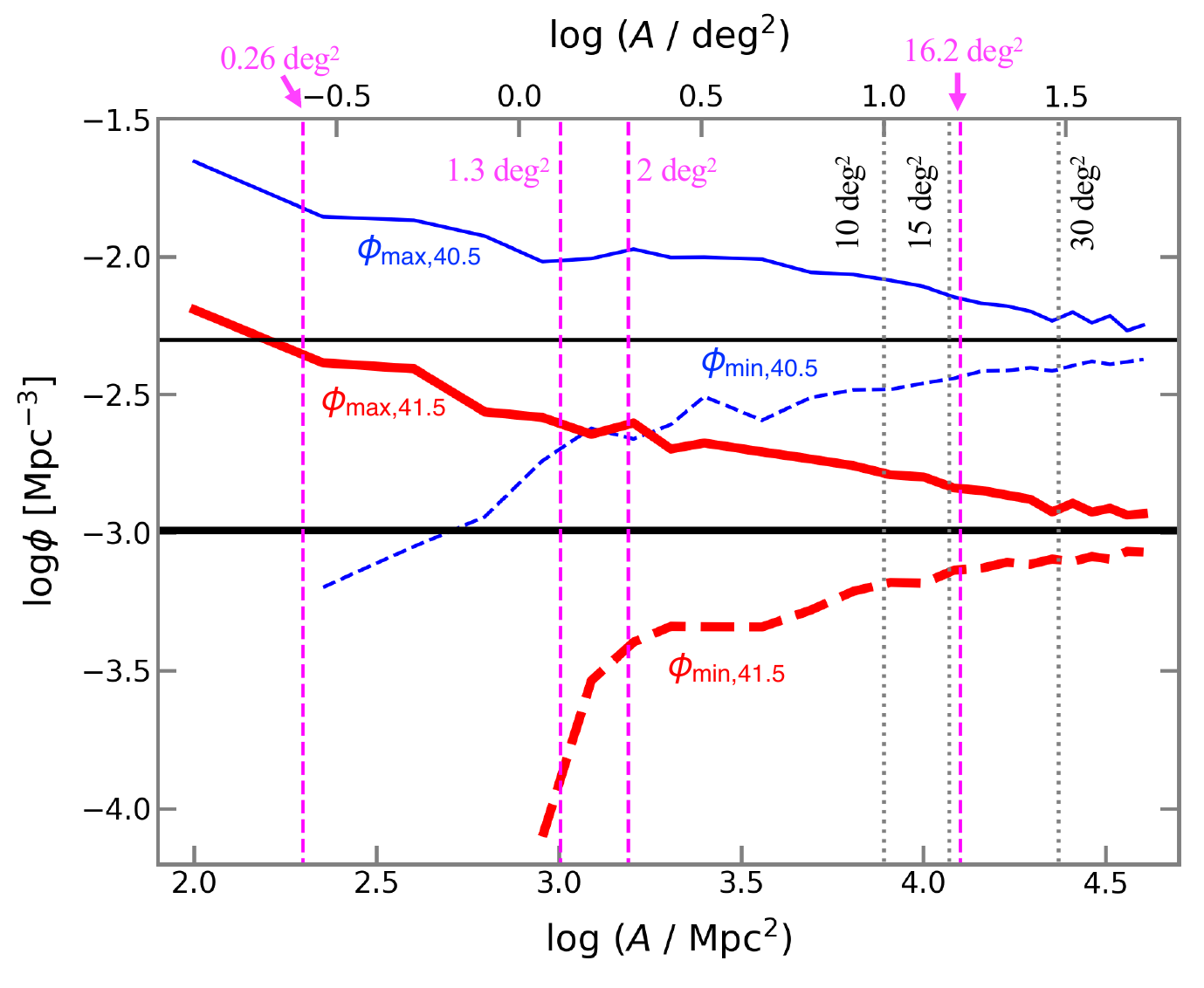}
\includegraphics[width=85mm]{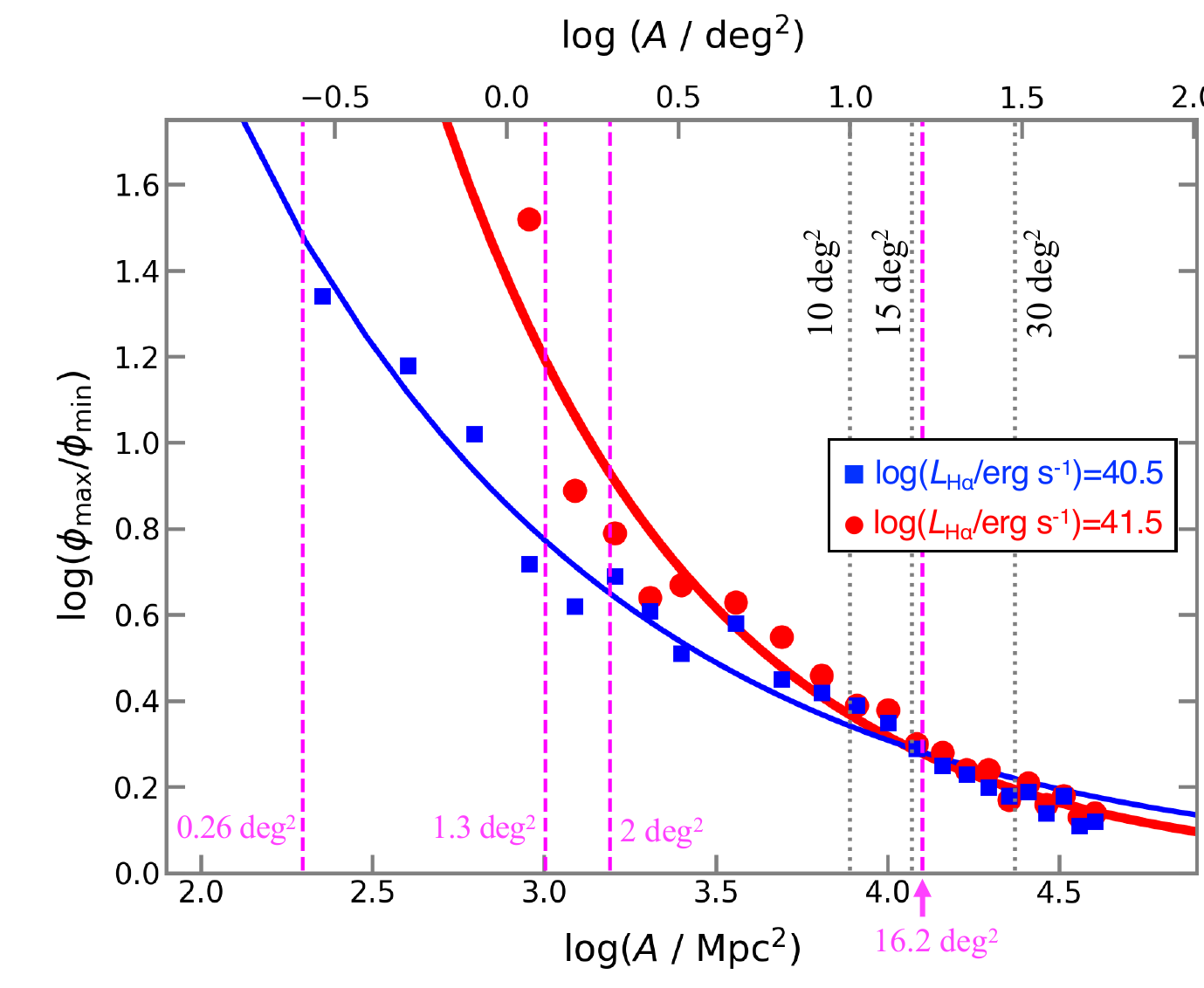}
\caption{
{\it Left}: The maximum (solid line) and minimum (dashed line) number density ($\phi$) of 
HAEs at H$\alpha$ luminosity of $L_{\rm H\alpha}=10^{40.5}$~erg~s$^{-1}$ (blue)
and at $L_{\rm H\alpha}=10^{41.5}$~erg~s$^{-1}$ (red). 
The maximum ($\phi_{\rm max}$) and ($\phi_{\rm min}$) number density within some survey
areas are summarized in Table \ref{tab:LF_MinMaxStdev}.
The black thin and thick horizontal lines indicate $\phi$ derived in whole of (411.8 Mpc)$^3$ 
box at $L_{\rm H\alpha}=10^{40.5}$~erg~s$^{-1}$ and $L_{\rm H\alpha}=10^{41.5}$~erg~s$^{-1}$,
respectively.
{\it Right}: The difference between the maximum logarithmic number density ($\phi_{\rm max}$) 
and minimum one ($\phi_{\rm min}$) at $L_{\rm H\alpha} = 10^{40.5}$~erg~s$^{-1}$ (blue)
and at $L_{\rm H\alpha} = 10^{41.5}$~erg~s$^{-1}$ (red).
Filled blue squares and red circles show log($\phi_{\rm max}$/$\phi_{\rm min}$) calculated in 
square regions by changing the length of a side from 10 Mpc to 200 Mpc at intervals of 10 Mpc. 
The blue thin and red thick solid curves show a power law fits: 
${\rm log}(\phi_{\rm max}/\phi_{\rm min})_{40.5} =  (12.20 \pm 1.40) \times A^{(-0.40 \pm 0.02)}$
at $L_{\rm H\alpha} = 10^{40.5}$~erg~s$^{-1}$
and ${\rm log}(\phi_{\rm max}/\phi_{\rm min})_{41.5} =  (63.80 \pm 20.8) \times A^{(-0.58 \pm 0.04)}$
at $L_{\rm H\alpha} = 10^{41.5}$~erg~s$^{-1}$.
In each panel, vertical dotted lines show corresponding survey area in deg$^2$
while magenta ones denote survey areas of \cite{2007ApJ...657..738L} [0.26~deg$^2$], 
\cite{2013MNRAS.433..796D} [1.3~deg$^2$], \cite{2011MNRAS.411..675S} [2~deg$^2$], 
and \cite{2018PASJ...70S..17H} [16.2~deg$^2$].
\label{fig:LF_MaxMin} 
}
\end{center}
\end{figure*}

In the left panel of Figure \ref{fig:LF_MaxMin}, we show the maximum ($\phi_{\rm max}$) and minimum 
($\phi_{\rm min}$) number density of HAEs at a H$\alpha$ luminosity of 
$L_{\rm H\alpha} = 10^{40.5}$~erg~s$^{-1}$ (blue lines) and $10^{41.5}$~erg s$^{-1}$ (red lines) 
as a function of survey areas. These are also summarized in Table \ref{tab:LF_MinMaxStdev}.
In the right panel, differences between the logarithmic $\phi_{\rm max}$ and 
$\phi_{\rm min}$ at $L_{\rm H\alpha} = 10^{40.5}$~erg~s$^{-1}$ (blue line and filled squares) and at 
$10^{41.5}$~erg~s$^{-1}$ (red line and filled circles) are shown.
As seen in these figures, the differences monotonically decrease with increasing area size. 
The difference at $L_{\rm H\alpha} = 10^{40.5}$~erg s$^{-1}$ (at $L_{\rm H\alpha} = 10^{41.5}$~erg~s$^{-1}$) 
is $\sim0.7$~dex ($\sim0.8$~dex) at the survey area of 2~deg$^2$ while it goes down to $\sim0.25$~dex 
($\sim0.3$~dex) in the case of 15~deg$^2$ survey area, for instance. 
The number density differences as a function of the survey area, $A$~[Mpc$^2$], are fitted by a power law 
function as follows: 
\begin{equation}
{\rm log}(\phi_{\rm max}/\phi_{\rm min})_{40.5} =  (12.20 \pm 1.40) \times A^{(-0.40 \pm 0.02)}
\end{equation}
at $L_{\rm H\alpha} = 10^{40.5}$~erg~s$^{-1}$ and
\begin{equation}
{\rm log}(\phi_{\rm max}/\phi_{\rm min})_{41.5} =  (63.80 \pm 20.8) \times A^{(-0.58 \pm 0.04)}
\end{equation}
at $L_{\rm H\alpha} = 10^{41.5}$~erg~s$^{-1}$.

\begin{figure} 
\begin{center}
\includegraphics[width=85mm]{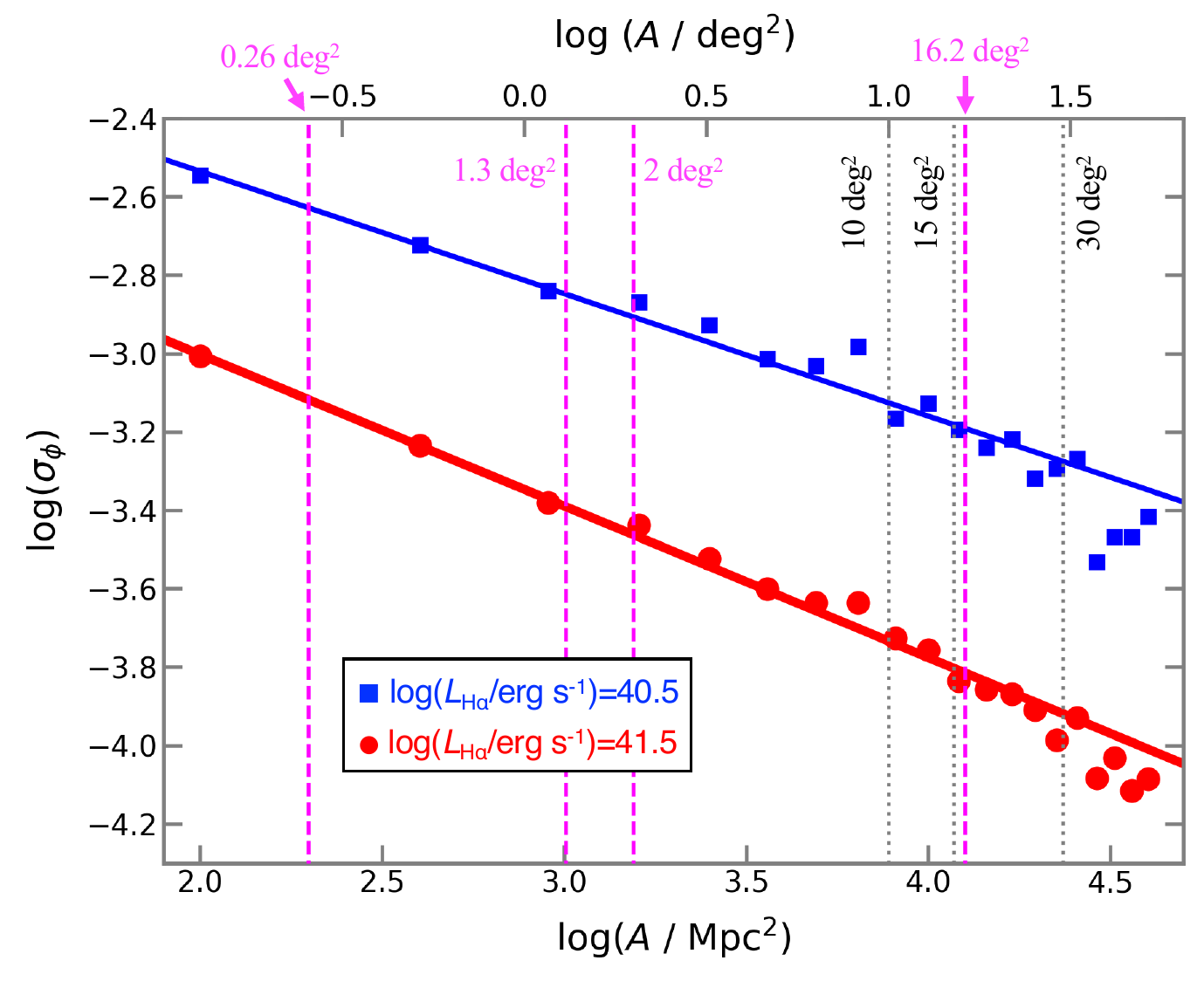}
\caption{The standard deviation of the number density of HAEs at $L_{\rm H\alpha}=10^{40.5}$ erg s$^{-1}$ 
(filled blue squares) and 10$^{41.5}$ erg s$^{-1}$ (filled red circles) as a function of survey areas
(the logarithm of the standard deviation of number density in the linear scale).
$\sigma_{\phi}$ shown by filled blue squares and red circles are calculated in square regions by changing the 
length of a side from 10 Mpc to 200 Mpc at intervals of 10 Mpc. 
The blue and red solid lines show power law fits:
$\sigma_{\phi,40.5} =  (1.20 \pm 0.10) \times10^{-2} \times A^{(-0.31 \pm 0.01)}$ at 
$L_{\rm H\alpha}=10^{40.5}$ erg s$^{-1}$ and 
$\sigma_{\phi,41.5} =  (5.79 \pm 0.21)\times10^{-3} \times A^{(-0.38 \pm 0.01)}$ at
$L_{\rm H\alpha}=10^{41.5}$ erg s$^{-1}$, respectively.
The vertical lines are the same as those in Figure \ref{fig:LF_MaxMin}.
\label{fig:LF_stdev} }
\end{center}
\end{figure}

Finally, we focus on the standard deviation of the number density of HAEs in our model.
Figure \ref{fig:LF_stdev} shows the standard deviation ($\sigma_{\phi}$) at $L_{\rm H\alpha} = 10^{40.5}$~erg~s$^{-1}$ 
(blue) and $10^{41.5}$~erg~s$^{-1}$ (red) as a function of $A$~[Mpc$^2$]. 
We summarize $\sigma_{\phi}$ within some survey areas in Table \ref{tab:LF_MinMaxStdev} (the third column).
$\sigma_{\phi}$ monotonically decreases with increasing area as well as the logarithmic $\phi$ difference.
The relation between $\sigma_{\phi}$ and $A$ [Mpc$^2$] can also be fitted by a power law:
\begin{equation}
\sigma_{\phi,40.5} =  (1.20 \pm 0.10) \times10^{-2} \times A^{(-0.31 \pm 0.01)}
\end{equation}
at $L_{\rm H\alpha} = 10^{40.5}$ erg s$^{-1}$ and
\begin{equation}
\sigma_{\phi,41.5} =  (5.79 \pm 0.21)\times10^{-3} \times A^{(-0.38 \pm 0.01)}
\end{equation}
at $L_{\rm H\alpha} = 10^{41.5}$ erg s$^{-1}$.
In the case of 10~deg$^2$ survey, $\sigma_{\phi,40.5} \sim8.0 \times 10^{-4}$ and $\sigma_{\phi,41.5} \sim1.8 \times 10^{-4}$ 
while $\sigma_{\phi,40.0} \sim1.3 \times 10^{-3}$ and $\sigma_{\phi,41.5} \sim3.6 \times 10^{-4}$ in a 2~deg$^2$ survey.

Based on these examination, it is suggested that the dispersion of observed H$\alpha$ LFs and the apparent agreement 
in observational results with different HAE selection criteria could be explained by field variance.
Equations (12) -- (15) can be used for estimating how the field variance could affect observed LFs for a given survey area.
Note that we have found that the H$\alpha$ LF shows a similar field variance even when we apply different set of parameters 
($f_{\rm neb}$, $EW_0$, and $f_{\rm gas}$) and thus the parameter values do not affect our discussion.

\begin{table}[h!]
\renewcommand{\thetable}{\arabic{table}}
\centering
\caption{Summary of the field variance of the H$\alpha$ LFs within various survey areas.} \label{tab:LF_MinMaxStdev}
\begin{tabular}{cccc}
\tablewidth{0pt}
\hline
\hline
\multicolumn{4}{c}{at $L_{\rm H\alpha}=10^{40.5}$ erg s$^{-1}$} \\
\hline
Area [deg$^2$] & $\phi_{\rm max}$ ($\times10^{-3}$) & $\phi_{\rm min}$ ($\times10^{-3}$) & $\sigma_{\phi}$ ($\times10^{-3}$) \\
\hline
\decimals
0.3 & 13.97 & 0.63 & 2.26\\
1.3 & 10.46 & 1.95 & 1.43 \\
2.0 & 10.43 & 1.50 & 1.33 \\
5.0 & 9.57 & 2.64 & 1.04 \\
10.0 & 8.12 & 3.26 & 0.80 \\
15.0 & 7.68 & 3.70 & 0.72 \\
30.0 & 5.92 & 3.88 & 0.45 \\
\hline
\hline
\multicolumn{4}{c}{at $L_{\rm H\alpha}=10^{41.5}$ erg s$^{-1}$} \\
\hline
Area [deg$^2$] & $\phi_{\rm max}$ ($\times10^{-3}$) & $\phi_{\rm min}$ ($\times10^{-3}$) & $\sigma_{\phi}$ ($\times10^{-3}$) \\
\hline
0.3 & 4.13 & 0.00 & 0.71 \\
1.3 & 2.44 & 0.21 & 0.40 \\
2.0 & 2.44 & 0.28 & 0.36 \\
5.0 & 1.95 & 0.46 & 0.25 \\
10.0 & 1.66 & 0.64 & 0.18 \\
15.0 & 1.53 & 0.71 & 0.16 \\
30.0 & 1.22 & 0.81 & 0.10 \\
\hline
\end{tabular}
\end{table}

\section{Concluding Remarks} \label{sec:conclud} 
We have constructed a semi-analytic model of HAEs at $z=0.4$ based on the $\nu^2$GC to examine the field variance of 
the HAE distribution. We define the HAE in our model to be a galaxy with H$\alpha$ $EW_0 \geq 40$ {\AA}. 
To calculate the observed H$\alpha$ luminosity, we assume the dust attenuation level of the H$\alpha$ emission is higher 
than that of the continuum emission, $f_{\rm neb}= 0.5$, as reported by observational studies at $z<1$.
As for the starburst galaxies, we assume that only galaxies having significant gas fraction ($f_{\rm gas} \geq 0.1$) emit 
H$\alpha$ emission.

This model well reproduces the observed H$\alpha$ LFs at similar redshift ($z=0.4$). The relationship between the SFR 
and stellar mass of HAEs is also reproduced by the model. HAEs seem to be typical MS galaxies because they show 
a tight correlation between the SFR and stellar mass;
${\rm log}[SFR / M_{\odot}~{\rm yr}^{-1}] = (0.86 \pm 0.02) {\rm log}[M_{\rm star} / M_{\odot}] - (8.26 \pm 0.24)$, 
with a scatter of $\sigma_{\rm MS} = 0.48\pm0.18$. 
In the model, the MS of HAEs is similar to that of galaxies with $sSFR \geq 10^{-10}$ yr$^{-1}$.

The (411.8 Mpc)$^3$ comoving volume of $\nu^2$GC enables us to examine the spatial distribution of galaxies over a wide 
area. The surface number density of HAEs with $L_{\rm H\alpha} \geq 10^{40}$ erg s$^{-1}$ is 308.9 deg$^{-2}$. 
We have found that the HAE is a good tracer of the large-scale structure in the Universe because their spatial distribution shows 
significant overdensity ($>5\sigma$) in cluster environments and cosmic web filaments. 

We have also examined the fluctuation of H$\alpha$ LFs derived in various survey areas. 
We have found that the H$\alpha$ LF derived by 2 deg$^2$ survey, typical area for previous observations, show significant field 
variance up to $\sim1$ dex. In the case of wider surveys of $\gtrsim$15 deg$^2$, the field variance of LFs becomes smaller
($\lesssim$0.3 dex) and converge to the average LF derived in the whole (411.8 Mpc)$^3$ box. 
The differences between maximum number density and minimum one can be fitted by a power law function,
as well as the standard deviation of the LF given by Equations (12) -- (15).
Based on these, one can estimate the variance of the LF to examine whether or not the survey area is enough for their
scientific goals.

This study shows that the $\nu^2$GC is so useful to examine various properties of galaxies over a wide area. 
We will construct HAE models at higher redshifts based on this study. Modeling other line emitters such as Ly$\alpha$ 
emitters, [O~{\sc ii}] emitters, and [O~{\sc iii}] emitters will be also done in future works.
To examine the consistency between the model and observational outcomes, precious measurements of the line luminosity
and EW are required. Further surveys with wide field spectroscopic instruments such as Prime Focus Spectrograph 
\citep{2014PASJ...66R...1T, 2016SPIE.9908E..1MT, 2018SPIE10702E..1CT} on the Subaru telescope are crucially important to 
progress our understandings on the nature of the galaxy evolution. 

\acknowledgments
We are very grateful to the anonymous referee for his/her careful reading, many useful comments and helpful suggestions.
MN, RS, KO, and MO are supported by grant from JSPS KAKENHI Nos 17H02867 (MN), 19K14766 (RS), 16K05299 (KO), 
and 17K14257 (MO). TI has been supported by MEXT as ``Priority Issue on Post-K computer'' (Elucidation of the Fundamental 
Laws and Evolution of the Universe), JICFuS and MEXT/JSPS KAKENHI Grant Number JP17H04828 (TI), JP17H01101 (TI), 
JP18H04337 (TI). 
We wish to recognize and acknowledge the very significant cultural role and reverence that the summit of Maunakea has always 
had within the indigenous Hawaiian community. We are most fortunate to have the opportunity to conduct observations from 
this mountain.

%




\end{document}